\documentclass[aps,prd,preprintnumbers,superscriptaddress,nofootinbib]{revtex4}
\usepackage{ulem}
\usepackage[dvips]{graphicx}
\usepackage{bm,latexsym,amsmath,amssymb,amsfonts}
\usepackage[colorlinks=true,linkcolor=magenta,citecolor=magenta]{hyperref}
\usepackage{color}
\input{colordvi.tex}

\newcommand*{\ba}{\begin{eqnarray}}
	\newcommand*{\ea}{\end{eqnarray}}

\newcommand{\simgt}{\lower.5ex\hbox{$\; \buildrel > \over \sim \;$}}
\newcommand{\simlt}{\lower.5ex\hbox{$\; \buildrel < \over \sim \;$}}
\newcommand*{\eps}{\varepsilon}
\newcommand*{\Lag}{{\cal L}}

\newcommand*{\p}{\partial}

\newcommand*{\tn}{\theta}
\newcommand*{\qd}{{\dot q}}
\newcommand*{\qdd}{{\ddot q}}
\newcommand*{\tnd}{{\dot \tn}}
\newcommand*{\tndd}{{\ddot \tn}}

\newcommand*{\tnh}{{\hat\theta}}
\newcommand*{\pih}{{\hat\pi}}

\newcommand*{\psib}{{\bar \psi}}
\newcommand*{\alphad}{{\dot \alpha}}
\newcommand*{\betad}{{\dot \beta}}
\newcommand*{\bx}{{\bf x}}
\newcommand*{\by}{{\bf y}}
\newcommand*{\bz}{{\bf z}}

\newcommand*{\B}{{\cal B}}
\newcommand*{\C}{{\cal C}}
\newcommand*{\E}{{\cal E}}
\newcommand*{\Y}{{\cal Y}}

\newcommand*{\F}{{\cal F}}
\newcommand*{\G}{{\cal G}}

\begin{document}

\title{Ghost free boson-fermion co-existence system}

\author{Rampei~Kimura}
\email[Email: ]{rampei"at"th.phys.titech.ac.jp}
\affiliation{Department of Physics, Tokyo Institute of Technology, Tokyo
152-8551, Japan}

\author{Yuki~Sakakihara}
\email[Email: ]{y.sakaki"at"sci.osaka-cu.ac.jp}
\affiliation{Department of Mathematics and Physics, Osaka City University, Osaka 558-8585, Japan}

\author{Masahide~Yamaguchi}
\email[Email: ]{gucci"at"phys.titech.ac.jp}
\affiliation{Department of Physics, Tokyo Institute of Technology, Tokyo
152-8551, Japan}

\begin{abstract}
We study co-existence system of both bosonic and fermionic degrees of freedom. Even if Lagrangian does not include higher derivatives, fermionic ghosts exist. For Lagrangian with up to first derivatives, we find the fermionic ghost-free condition in Hamiltonian analysis, which is found to be the same with requiring that the equations of motion of fermions are first-order in Lagrangian formulation. When fermionic degrees of freedom are present, uniqueness of time evolution is not guaranteed a priori because of the Grassmann property.
We confirm that the additional condition, which is introduced to close Hamiltonian analysis, also ensures the uniqueness of the time evolution of system.
\end{abstract}

\preprint{OCU-PHYS-462, AP-GR-137}
\maketitle
\section{Introduction}

The presence of inflation and dark energy, the past and the current
acceleration of the Universe, is strongly supported by recent
observations, e.g., of cosmic microwave background
anisotropies~\cite{Ade:2015xua,Ade:2015lrj} and of 
supernovae~\cite{Perlmutter:1998hx,Riess:1998cb}. However, we have not
yet identified what caused inflation and what causes the current
acceleration of the Universe. If we had the unique ultimate
theory, it would automatically predict the past and current acceleration and one
could easily identify the fields responsible for them. Unfortunately we 
have not yet discovered such a theory, and hence, we have to pin down the true
theory step by step through observational results. For such a purpose, it will be 
quite useful to consider a general theory realizing inflation and/or dark
energy because the true theory would lie in such a framework if it is
wide enough.

One of the famous examples of such general theories is Horndeski
theory~\cite{Horndeski:1974wa}, which is the most general (single-field)
scalar tensor theory with second-order equations of motion to avoid
ghost instabilities. This theory, originally proposed by Horndeski
more than 40 years ago, was recently
rediscovered~\cite{Deffayet:2011gz} in the context of Galileon
theory~\cite{Nicolis:2008in}, and their equivalence was proven in
Ref.~\cite{Kobayashi:2011nu}. It was, however,
noticed~\cite{Gleyzes:2014dya,Gleyzes:2014qga,Zumalacarregui:2013pma}
that the requirement of the second-order nature of Euler-Lagrange equations
is more than enough to avoid ghost instabilities. A wider class of
models~\cite{Langlois:2015cwa,Langlois:2015skt,Crisostomi:2016tcp,Crisostomi:2016czh,Achour:2016rkg,BenAchour:2016fzp,Gao:2014soa,Lin:2014jga,Gao:2014fra}
can realize healthy scalar-tensor theories without the ghost instabilities
associated with higher derivatives.  Another interesting example to
consider a generic theory is the effective field theory approach to
inflation~\cite{Weinberg:2008hq,Cheung:2007st} and to dark
energy~\cite{Gubitosi:2012hu,Bloomfield:2012ff,Gleyzes:2013ooa}. In
\cite{Cheung:2007st}, cosmological perturbations are controlled by the
symmetry of the background cosmology, and any terms respecting this
symmetry can appear. This approach is in some sense wider than the previous
one in that it can accommodate higher derivative terms leading to ghosts as long as
the cutoff scale around which the ghosts would appear is above
the scale we are interested in. On the other hand, it straightforwardly
implies that such higher derivative terms cannot play a dominant role of
the dynamics because, otherwise, the associated ghosts would appear even at
the scale we are interested in. Thus, though both approaches are
complementary and have their pros and cons respectively, we confine
ourselves to the former approach and search for a general theory free from ghost instabilities in this paper.

The former approach was recently extended even to vector-tensor theory
\cite{Deffayet:2013tca,Heisenberg:2014rta,Allys:2015sht,Jimenez:2016isa,Heisenberg:2016eld,Fleury:2014qfa,Allys:2016jaq,Kimura:2016rzw}.
However, as far as we know,
no one has tried to extend this approach (in fact, both approaches) so as
to include fermionic degrees of freedom. Then we are led to a question of what is the most general theory including fermionic degrees of freedom
without ghost instabilities.
As discussed in \cite{Ribas:2005vr,Ribas:2007qm,deSouza:2008az}, fermionic matter can be responsible for the acceleration of the universe. Even if fermionic degrees of freedom do not dominate the universe, the whole Lagrangian should include fermions as standard model particles, whose effect might be observed through the loop corrections to the bispectrum of primordial curvature perturbations as pointed out in \cite{Chen:2016nrs,Chen:2016uwp}. Their couplings to the inflaton are also important when one discusses the reheating stage of inflation.
A generic discussion in the context of the effective field theory approach to reheating is given in Refs.~\cite{Ozsoy:2015rna,Giblin:2017qjp}, though it is confined to bosonic degrees of freedom.  
In this paper, according to the same spirit with
Ref.~\cite{Motohashi:2016ftl}
(see also \cite{Klein:2016aiq} for a complementary analysis and 
\cite{Langlois:2015cwa,Crisostomi:2016czh,BenAchour:2016fzp,Crisostomi:2017aim}
for field theoretical extensions.), as a first step, we begin with
point particle theory with both bosonic and fermionic degrees of freedom
and derive the ghost-free condition. The extension to field theory and
higher derivatives will be given in a further publication soon.

This paper is organized as follows. In the next section, we review the properties of Grassmann algebra and fermions based on the
textbook written by Henneaux and Teitelboim~\cite{Henneaux:1992ig}. 
In Sec.~\ref{section3}, we concentrate on the purely fermionic system and explain that the
absence of negative norm states requires the reduction in the dimension of phase space.  
In Sec.~\ref{section4}, 
we give our setup of the coexistence system with bosonic and
fermionic point particles and derive the condition for avoiding
fermionic ghosts, which we call the maximally-degenerate condition. We then perform a Hamiltonian analysis of the
system satisfying the condition. We 
find another condition guaranteeing that secondary constraints are not produced and all the Lagrange multipliers are uniquely determined. 
In the last part of Sec.~\ref{section4},
we also show how these conditions can be understood in Lagrangian formulation.  
In Sec.~\ref{section5}, we provide concrete examples,
which are free from  fermionic ghosts, and explicitly show
the consistency with our analysis.
The final section is devoted to summary. 
In Appendix~\ref{AppendixB}, we mention how to produce primary constraints properly even when the maximally-degenerate condition is not satisfied. In Appendix~\ref{Appendix3}, we discuss Hamiltonian analysis including fermions when we also possibly have secondary constraints.
In Appendix~\ref{Appendix2}, we explicitly prove that the maximally-degenerate condition is equivalent to the presence of $N$ primary constraints. 
In Appendix~\ref{Appendix4}, we calculate the Dirac brackets between canonical variables in the maximally-degenerate case.
In Appendix~\ref{Appendix5}, we give a simple extension to the ghost free boson-fermion co-existence field theory.

\section{Grassmann algebra and canonical formalism}

In quantum field theories, fermionic fields obey canonical
anti-commutation relations, $\{\psi_a(t, {\bf x}), \pi_b(t, {\bf
	y})\}_{+} =i\delta_{ab}({\bf x}-{\bf y})$, where $\psi_a$ is a fermion and
$\pi_a$ is its conjugate momentum.  For the purpose of constructing
a general action with bosons and fermions, we would like to start with the 
classical (or ``pseudo-classical'') treatment of them.  To deal with
fermions in the context of classical mechanics, 
one needs to reformulate canonical formalism such that classical analysis is 
consistent with anti-commutation relations in quantum theory.
In the first part of this section, we briefly provide an overview of the
basics of Grassmann algebra.  Then, we focus on Hamiltonian formulation
including fermionic degrees of freedom.  (All the materials described in
this section and Sec.~\ref{section3} are based on
\cite{Henneaux:1992ig}.)

\subsection{Grassmann algebra}

A Grassmann algebra is formed by generators $\xi^A$ with
$A=1, 2, ..., M$ satisfying the anti-symmetric relations, $\xi^A \xi^B + \xi^B \xi^A
=0$. From this definition, it is clear that each generator squared
should be zero, $\xi^A \xi^A =0$ (no summation), which suggests the Pauli
exclusion principle at the level of classical mechanics.  In terms of
generators $\xi^A$, an arbitrary function $g$ can be expressed as
\ba
g=g_0 + g_A \xi^A + g_{AB} \xi^A \xi^B + \cdots + g_{A_1...A_M} \xi^{A_1} ... \,\xi^{A_M} \ ,
\ea 
where the coefficients $g_{A_1...A_n} $ are completely anti-symmetric.
The terms made of an even (odd) number of $\xi^A$ are called ``Grassmann-even'' (``Grassmann-odd''). Now we introduce even dynamical variables $q^i(t)$
$(i=1,2,\cdots n)$ and odd ones $\tn^\alpha(t)$ $(\alpha=1,2,\cdots N)$ as follows,
\ba
q^i(t)&=&q_0^i (t) + q_{AB}^i (t) \xi^A \xi^B + \cdots \ , \\
\tn^\alpha(t)&=&\theta_A^\alpha (t) \xi^A+ \tn_{ABC}^\alpha (t) \xi^A \xi^B \xi^C+ \cdots \ ,
\ea
where the coefficients $q_{A_1...A_n}^i$ and $\tn_{A_1...A_n}^\alpha$
are completely anti-symmetric and time-dependent. 
(Since we do not require the covariance, superscripts and subscripts are just labels of the variables, e.g., $\theta^\alpha=\theta_\alpha$, except for Appendix \ref{Appendix5}.)
These variables satisfy
the following (anti-)commutation relations:
\ba
&& q^i q^j -q^j q^i = 0 \ , \\
&& \tn^\alpha q^i -q^i \tn^\alpha =0 \ ,\\ 
&& \tn^\alpha \tn^\beta +\tn^\beta \tn^\alpha =0 \ . 
\ea
From the above relations, $q^i(t)$ can be regarded as bosons and
$\tn^\alpha(t)$ as fermions.

\begin{description}
\item[Function] Because of Grassmann nature, an arbitrary (super)function $f(q^i, \tn^\alpha)$,
which depends on $\xi^A$ only through $q^i$ and $\tn^\alpha$,
can be expanded in powers of the odd variables $\tn^\alpha$, 
\ba
f(q^i,~\tn^\alpha)= f_0(q^i) + f_\alpha(q^i)  \tn^\alpha +  f_{\alpha\beta}(q^i)  \tn^\alpha \tn^\beta + \cdots \ , 
\label{EXsuper}
\ea
where $f_0(q^i)$ and $f_{\alpha_1...\alpha_k}(q^i)$ are Grassmann-even functions
with fully anti-symmetric indices. 

\item[Derivative] Left derivatives with respect to Grassmann-odd variables are defined by removing the variable from the left, 
\ba
\delta f = \delta \tn^\alpha \frac{\partial^L f}{\partial \tn^\alpha} \ .
\ea
Throughout this paper, we use left derivatives and omit the superscript $L$ for the derivative operator, $\partial/\partial \tn^\alpha \equiv \partial^L/\partial \tn^\alpha$.

\item[Complex conjugate] Let us define the complex conjugate in Grassmann algebra.
In order to be consistent with Hermitian conjugation of operators, 
the complex conjugation is required to have the following properties: 
\ba
(\tn^\alpha \tn^\beta)^* &=& \tn^{\beta\,*} \,\tn^{\alpha\,*} ,\\
(\tn^{\alpha\,*})^* &=& \tn^\alpha ,\\
(a \, \tn^\alpha)^* &=& a^* \,\tn^{\alpha\,*} ,
\label{prop_Grassmann}
\ea
where $a$ is a complex number.

\item[Inverse matrix] Whether a matrix is invertible or not plays an important role in degenerate theories as we will see in Sec.~IV. The condition for the existence of the inverse matrix of a Grassmann valued square matrix is obtained as follows. We introduce two square matrices that are functions of the variables $q^i$ and $\tn^\alpha$, which can be in general written as 
\ba
A(q^i, \tn^\alpha) &=&  A_0(q^i) + A_\alpha(q^i)  \tn^\alpha + A_{\alpha\beta}(q^i) \tn^\alpha \tn^\beta+\cdots \ ,\\
B(q^i, \tn^\alpha) &=&  B_0(q^i) + B_\alpha(q^i)  \tn^\alpha + B_{\alpha\beta}(q^i) \tn^\alpha \tn^\beta+\cdots \ ,
\ea
where $A_0, A_\alpha, ..., B_0, B_\alpha, ...$ are fully anti-symmetric matrices depending on $q^i$. 
The condition that $B$ be the inverse of $A$ is given by $AB=I$, where $I$ is the identity matrix, 
which leads to the following equations, 
\ba
A_0 B_0 &=& I \ , \\
A_0 B_\alpha + A_\alpha B_0 &=&0 \ ,\\
A_0 B_{\alpha\beta}+{1 \over 2} (A_\alpha B_\beta-A_\beta B_\alpha) + A_{\alpha\beta}B_0 &=& 0 \ ,\\
&\vdots& \ .\nonumber
\ea
You will find that if and only if $A_0$ has the inverse, the equations can be solved successively as
\ba
B_0&=&A_0^{-1} \ , \\
B_\alpha &=& - A_0^{-1} A_\alpha A_0^{-1} \ , \\
B_{\alpha\beta} &=& {1\over 2} A_0^{-1} (A_\alpha A_0^{-1} A_\beta A_0^{-1} -A_\beta A_0^{-1} A_\alpha A_0^{-1})- A_0^{-1} A_{\alpha\beta} A_0^{-1} \ ,\\
&\vdots&\nonumber
\ea
which also satisfy $BA=I$.
Therefore, we conclude that {\it a matrix $A$ has the inverse if and only if $A_0$ has the inverse,} i.e.,
\ba
\det (A_0) \neq 0 \ , \qquad {\rm where} \quad A_0=A|_{\tn = 0} \ .
\label{conditionInv}
\ea

\end{description}

\subsection{Hamiltonian formulation}

Now we move on to Hamiltonian formulation both with $n$ Grassmann-even variables $q^i(t)$ and $N$ Grassmann-odd ones $\theta^\alpha(t)$. In the present paper, we consider the Lagrangian containing up to the first derivatives, namely,
\ba
S = \int^{t_2}_{t_1} L(q^i, \qd^i, \tn^\alpha, \tnd^\alpha) dt \ .
\label{Lagrangian}
\ea
We require that the Lagrangian be an even and real function, since the Hamiltonian is the generator of time evolution. The dynamical variables $q^i$ and $\tn^\alpha$ are taken to be real\,\footnote{Since complex variables can be decomposed into real and imaginary parts and expressed in terms of a set of two real variables, we can always identify complex variables with doubled real variables without loss of generality.} 
throughout this paper.
The variations with respect to $z^A = (q^i, \tn^\alpha)$ yield Euler-Lagrange equations,
\ba
\frac{d}{dt} \left( \frac{\p L}{\p {\dot z}^A}\right)-\frac{\p L}{\p  z^A}=0 \ ,
\label{ELeqn}
\ea
where we require the variations to vanish at the endpoints, $\delta z_A (t_1)=\delta z_A (t_2)=0$.
The canonical momenta are defined by 
\ba
p_i = \frac{\p L}{\p \qd^i} \ , \quad \pi_\alpha =\frac{\p L}{\p \tnd^\alpha} \ . 
\label{momenta}
\ea
Note that $p_i$ are even and real variables as usual, while $\pi_\alpha$ are odd and imaginary variables\,\footnote{Please refer to footnote~6 for the details.}, since the Lagrangian is real and even.
Then, the Hamiltonian is given by
\ba
H= \qd^i p_i + \tnd^\alpha \pi_\alpha 
-L(q^i,\qd^i, \tn^\alpha, \tnd^\alpha) \,
\label{Hamiltonian}
\ea
and the variational principle of action yields Hamilton's equations,
\ba
\qd^i = \frac{\p H}{\p p_i} \ , \quad {\dot p}_i = -\frac{\p H}{\p q^i} \ , \qquad 
\tnd^\alpha =-\frac{\p H}{\p \pi_\alpha} \ , \quad {\dot \pi}_\alpha = -\frac{\p H}{\p \tn^\alpha} \ .
\label{HamiltonEq}
\ea
It should be noticed that the minus sign appears in the third equation.
The time-evolution of a function $F(q^i, \tn^\alpha)$ can be expressed as 
${\dot F} = \partial F/\partial t+ \{F, H\}$, 
where the Poisson bracket between arbitrary functions $F$ and $G$ is~\cite{1976NCimA..33..115C}
\ba
\{F, G\} = \left(\frac{\p F}{\p q^i}\frac{\p G}{\p p_i}-\frac{\p F}{\p p_i}\frac{\p G}{\p q^i}\right)
+(-)^{\eps_F}
\left(\frac{\p F}{\p \tn^\alpha}\frac{\p G}{\p \pi_\alpha}+\frac{\p F}{\p \pi_\alpha}\frac{\p G}{\p \tn^\alpha}\right) \ .
\label{PB}
\ea
Here, $\eps_F$ represents the Grassmann parity of $F$, i.e., $\eps_F=0$ if $F$ is even, and $\eps_F=1$ if $F$ is odd.
As a consequence, the Poisson brackets between the canonical variables are found to be
\ba
&&\{q^i, p_j\} = \delta^i_{~j} \ , \quad \{\theta^\alpha, \pi_\beta\} = - \delta^\alpha_{~\beta} \ , \\
&&\{q^i, q^j\} = \{p_i, p_j\} =\{\theta^\alpha, \theta^\beta\} =\{\pi_\alpha, \pi_\beta\}=0 \ .
\ea
The Poisson bracket satisfies the following identities,
\ba
\{F,~G\} &=&(-)^{\eps_F \eps_G+1} \{G,~F\} \ , \label{P1}\\
\{F,~G_1G_2\} &=&\{F,~G_1\}G_2 + (-)^{\eps_F \eps_{G_1}} G_1 \{F,~G_2\} \ , \label{P2}\\
\{F_1F_2,~G\} &=&F_1\{F_2,~G\} + (-)^{\eps_{F_2} \eps_{G}} \{F_1,~G\} F_2 \ , \label{P3}
\ea
which are easily proved from the definition of the Poisson bracket~(\ref{PB}).

The prescription of the canonical quantization is simply
replacing the Poisson brackets between canonical variables by commutation relations for bosons and by anti-commutation relations for fermions as 
\ba
 \{A,B\} \to 
\begin{cases}
	(i \hbar )^{-1} \{{\hat A},{\hat B}\}_{-} \quad & \textrm{if $A$ and $B$ are bosons}, \\
	(i \hbar )^{-1} \{{\hat A},{\hat B}\}_{-} \quad & \textrm{if $A$ is a boson (fermion) and $B$ is a fermion (boson)}, \\
	(i \hbar )^{-1} \{{\hat A},{\hat B}\}_{+} \quad & \textrm{if $A$ and $B$ are fermions},
\end{cases}
\label{CQ}
\ea
where the commutator and anti-commutator are respectively defined as $\{{\hat A},{\hat B}\}_{-} = {\hat A}{\hat B}-{\hat B}{\hat A}$ and $\{{\hat A},{\hat B}\}_{+}= {\hat A}{\hat B}+{\hat B}{\hat A}$. 
If the system contains (second class) constraints\,\footnote{When the system contains first class constraints (sometimes in addition to second class constraints), we can add gauge fixing conditions to the set of the constraints, which effectively leads to a system only with second class constraints.},
one should rather use the Dirac bracket instead of the Poisson bracket, defined by
\ba
\{A,B\}_D=\{A,B\}-\{A,\phi_a\}(C^{-1})^{ab}\{\phi_b,B\} \ ,
\ea
where $\phi_a$ are second class constraints and $C_{ab}=\{\phi_a,\phi_b\}$. 
Hereafter, we set $\hbar=1$ in this paper.
One should note that real variables such as $q^i, p_i$, and $\tn^\alpha$ will be promoted to Hermitian operators, and the imaginary variables $\pi_\alpha$ then become anti-Hermitian operators through the canonical quantization.

\section{Necessity of degeneracy - Example: purely fermionic system}
\label{section3}

Contrary to a purely bosonic system, an $N$-fermionic system needs constraints eliminating $N/2$ ghostly degrees of freedom to realize a healthy system. We will see the appearance of negative norm states for the fermionic system without any constraints, i.e., in non-degenerate theories. We also show how the negative norm states are avoided for the usual Weyl-type fermions. 
We omit bosonic degrees of freedom here for simplicity, but the essence does not change for the boson-fermion co-existence system as we will see in the next section.

\subsection{Non-degenerate fermionic system}

In this subsection, we begin with the action given by
\ba
S = \int^{t_2}_{t_1} L(\tn^\alpha, \tnd^\alpha) dt \ .
\label{LagrangianF}
\ea
We assume that the Lagrangian is non-degenerate, 
\begin{align}
 \det \left(\frac{\partial^2 L}{\partial \dot{\theta}^\beta\partial\dot{\theta}^\alpha}\right)^{(0)}\neq 0 \ , \qquad {\rm where} \quad 
\left(\frac{\partial^2 L}{\partial \dot{\theta}^\beta\partial\dot{\theta}^\alpha}\right)^{(0)}=\left.\frac{\partial^2 L}{\partial \dot{\theta}^\beta\partial\dot{\theta}^\alpha}\right|_{\theta,\dot{\theta}=0} \ ,
\end{align}
and the Euler-Lagrange equations~(\ref{ELeqn}) then contain the second time derivatives of $\tn^\alpha$.
In other words, this system does not have any constraints, and the total
number of degrees of freedom is the same as the number of the original variables $N$. (The phase space is spanned by $2N$ canonical variables.)

Now we would like to show that non-degenerate fermionic system inevitably gives
negative norm states.
Similar situations are known to be found in the non-degenerate Lagrangian with higher derivatives of bosonic variables. In the bosonic case, 
after the replacement of the higher derivative terms with newly defined
variables, one finds that the Hamiltonian is linear in momentum and not
bounded from below, which leads to Ostrogradsky's ghost
instability~\cite{Ostrogradsky:1850fid,Woodard:2006nt}. This ghost
can be interpreted as the appearance of negative norm states in the
quantized theory~\cite{Woodard:2015zca}. In a fermionic system, the
positivity of the Hamiltonian is not guaranteed at the classical level, and
we should discuss the stability after the quantization.
Based on the canonical quantization~(\ref{CQ}), we obtain anti-commutation relations,
\begin{align}
\begin{aligned}
&\{{\hat \tn}^\alpha, {\hat \pi}_\beta\}_+ = - i \delta^\alpha_{~\beta} \ , \\
&\{{\hat \tn}^\alpha, {\hat \theta}^\beta\}_+ =\{{\hat \pi}_\alpha, {\hat \pi}_\beta\}_+=0 \ . 
\label{ACR}
\end{aligned}
\end{align}
Here, the canonical operators ${\hat \tn}^\alpha$ and ${\hat \pi}_\alpha$ are now Hermitian and anti-Hermitian operators, respectively.
Then, we introduce orthogonal Hermitian operators,
\ba
{\hat A}_\alpha = {1 \over \sqrt{2}}({\hat \tn}_\alpha - i {\hat \pi}_\alpha) \ , \quad 
{\hat B}_\alpha = {1 \over \sqrt{2}}({\hat \tn}_\alpha + i {\hat \pi}_\alpha) \ ,
\ea
and the anti-commutation relations between them are given by 
\ba
\{{\hat A}_\alpha,{\hat A}_\beta\}_+ = - \delta_{\alpha\beta} \ , \quad
\{{\hat A}_\alpha,{\hat B}_\beta\}_+ = 0 \ , \quad
\{{\hat B}_\alpha,{\hat B}_\beta\}_+ =  \delta_{\alpha\beta} \ .
\ea
One immediately notices that all eigenvalues of the first anti-commutator have the negative sign, leading to the negative norm states,
while those of the third anti-commutator have the correct sign guaranteeing the positivity of the norm of states.
This fact tells us that 
{\it each fermionic degree of freedom in physical space should carry 1
degree of freedom in phase space, otherwise negative norm states
inevitably appear,} which implies that $N/2$ physical degrees of freedom ($N$ degrees of freedom in phase space) are extra degrees of freedom corresponding to fermionic ghosts. 
Since this is a direct consequence of the canonical
quantization of the canonical
variables $\tn^\alpha$ and $\pi_\alpha$~(\ref{ACR}), any fermionic non-degenerate
theories always suffer from negative norm states even if we have bosonic variables in addition.

\subsection{Degenerate fermionic system}

Although the appearance of negative norm states
seems to be a generic feature
 of a non-degenerate fermionic system as we saw in the previous subsection, 
we already known that a Weyl field, for example, does not suffer from 
such a problem.
Here, we review why such a theory can avoid negative norm states 
by illustrating a simple model, 
\ba
L= -{i \over 2} \tnd^\alpha \tn_\alpha \ .
\ea
Obviously, the Euler-Lagrange equations are first-order differential equations, 
and this model could be regarded as a classical counterpart of a Weyl fermion.
The canonical momenta are given by $\pi_\alpha = -i\tn_\alpha/2$, which lead to the primary constraints,
\ba
\phi_\alpha \equiv \pi_\alpha + {i \over 2} \tn_\alpha = 0 \ . 
\label{constraintDirac}
\ea
Since the Hamiltonian vanishes, $H=0$, 
the total Hamiltonian is simply given by $H_T=\phi_\alpha \mu^\alpha$, 
where $\mu^\alpha$ are the Lagrange multipliers\,\footnote{
The order of the constraints $\phi_\alpha$ and the Lagrange multipliers $\mu^\alpha$
in the total Hamiltonian should be like $\phi_\alpha \mu^\alpha$ in the left derivative notation. We also note that $\mu^\alpha$ are Grassmann-odd numbers.}.
The Poisson brackets between the primary constraints
are $\{\phi_\alpha, \phi_\beta\} = -i \delta_{\alpha\beta}$, which means that  
all $\phi_\alpha$ are second class constraints, and no further constraints are added. Then the time evolution of the constraints~(\ref{constraintDirac}) 
determines the Lagrange multipliers as
${\dot \phi}_\alpha = \{\phi_\alpha, \phi_\beta\} \mu^\beta =-i \mu_{\alpha} \approx0$, where $\approx$ means the weak equality. 
The dimension of the phase space spanned by the canonical variables is $2N$. Since we have $N$ (second class) primary constraints, the number of physical degrees of freedom is $(2N- N)/2=N/2$ as it should be.

For confirmation, we now check the absence of negative norm states in this system.
Since we have second class constraints, we evaluate the Dirac brackets between all canonical variables, 
\ba
\{\theta_\alpha, \theta_\beta\}_D = - i\delta_{\alpha\beta} \ ,\quad
\{\theta_\alpha, \pi_\beta\}_D= - {1\over 2} \delta_{\alpha\beta} \ , \quad
\{\pi_\alpha, \pi_\beta\}_D={i \over 4} \delta_{\alpha\beta} \ ,
\ea
and the canonical quantization leads to the following anti-commutation relations,
\ba
\{\tnh_\alpha, \tnh_\beta\}_+ = \delta_{\alpha\beta} \ ,\quad
\{\tnh_\alpha, \pih_\beta\}_+= - {i\over 2} \delta_{\alpha\beta} \ , \quad
\{\pih_\alpha, \pih_\beta\}_+=-{1 \over 4} \delta_{\alpha\beta} \ .
\label{CQinDirac}
\ea
One should note that these anti-commutation relations between the canonical variables are consistent with the primary constraints, i.e., plugging $\pih_\alpha =- i \tnh_\alpha /2$ into the second and the third expressions in (\ref{CQinDirac}) recovers the first one. 
It is clear that negative norm states do not appear in this system since the relations $\{\tnh_\alpha, \tnh_\beta\}_+$
are positive definite.\,\footnote{Here, we adopt $\tnh_\alpha$ as independent variables since they are Hermitian operators. If one would like to adopt $\pih_\alpha$ instead, they should be multiplied by $i$ to be Hermitian.}

\section{Degenerate theories in boson-fermion co-existence system}
\label{section4}
As seen in the previous section, the unique solution to avoid negative norm states in $N$-fermionic system is to have a sufficient number of constraints eliminating $N/2$ ghostly degrees of freedom. In this section, we provide a general approach to constructing a degenerate Lagrangian of the boson-fermion co-existence system, 
whose physical degrees of freedom are $n + N/2$ with $n$  the number of bosonic variables. 
We focus on the most general Lagrangian containing up to first time derivatives of bosons and fermions~(\ref{Lagrangian}). 
In the former part of this section, we derive a (sufficient) condition which yields $N$ constraints to eliminate fermionic  ghosts
in the Hamiltonian formulation. 
In the latter part, we show that 
the condition, imposed in Hamiltonian formulation, is equivalent to requiring that the equations of motion of fermions are first-order differential equations. 
We also introduce another condition, which we call the uniqueness condition, to have no more constraints in Hamiltonian formulation and show that it is responsible for the unique time evolution of the system in Lagrangian formulation.

\subsection{Degeneracy condition}
\label{Subsection:DegeneracyConditions}

If the time derivatives of $q^i$ and $\theta^\alpha$ are expressed in terms of the canonical variables $(q^i, p^i, \theta^\alpha, \pi^\alpha)$, we do not have any primary constraints. Therefore, we need to look for the condition where the time derivatives of $q^i$ and $\theta^\alpha$ are not written in terms of the canonical variables.
Let us then consider the infinitesimal variations of the canonical momenta
with respect to all variables,
\ba
\begin{pmatrix}
 \delta p_i \\
 \delta \pi_\alpha
\end{pmatrix}
=
K
\left(
\begin{array}{c}
	\delta \qd^j  \\
	\delta \tnd^\beta  \\
\end{array}
\right)+
\begin{pmatrix}
 L_{\dot{q}^i q^j}& -L_{\dot{q}^i \theta^\beta}\\
 L_{\dot{\theta}^\alpha q^j} & L_{\dot{\theta}^\alpha \theta^\beta}
\end{pmatrix}
\begin{pmatrix}
 \delta q^j \\
 \delta \theta^\beta
\end{pmatrix}\ ,
\label{infmom}
\ea
where $K$ is the kinetic matrix,
\ba
K=
\left(
\begin{array}{cc}
	A_{ij} & {\cal B}_{i\beta} \\
	{\cal C}_{\alpha j} & D_{\alpha\beta} \\
\end{array}
\right) \ ,
\ea
whose
components are defined by
\begin{align}
A_{ij} & =\frac{\partial p_i}{\partial \dot{q}^j}=L_{\qd^i \qd^j}  \ , \qquad
{\cal B}_{i\beta} =-\frac{\partial p_i}{\partial \dot{\theta}^\beta} =
 -L_{\qd^i \tnd^\beta} \ ,  \nonumber\\
{\cal C}_{\alpha j} &= \frac{\partial \pi_\alpha}{\partial \dot{q}^j}=
 L_{\tnd^\alpha \qd^j} \ , \qquad
D_{\alpha\beta} = \frac{\partial \pi_\alpha}{\partial \dot{\theta}^\beta}= L_{\tnd^\alpha \tnd^\beta} \,\left(
= - L_{\tnd^\beta \tnd^\alpha}
 \right)\ .
\end{align}
Here we have introduced an abbreviated notation, 
\begin{align}
 L_{XY}=\frac{\partial^2 L}{\partial Y\partial X}=\frac{\partial }{\partial Y}\Bigl(\frac{\partial L}{\partial X}\Bigr)\ .
\end{align} 
It should be noticed that all the sub-matrices depend on $(q^i,\dot{q}^i,\theta^\alpha,\dot{\theta}^\beta)$ in general. By construction, $A_{ij} $ is a Hermitian symmetric matrix, while $D_{\alpha\beta}$ is an anti-Hermitian anti-symmetric matrix, both of which are Grassmann-even\,\footnote{The product of two real fermionic variables is not real but imaginary because of the Grassmann property~\eqref{prop_Grassmann} and should always be accompanied by the imaginary unit $i$ in (Grassmann-even real) Lagrangian. Then, the matrix $D_{\alpha\beta}$ is a pure imaginary matrix, which is consistent with its anti-Hermitian property. For instance, when the Lagrangian includes $\frac{1}{2}\theta_\alpha\theta_\beta \dot{\theta}_\alpha\dot{\theta}_\beta$, $D_{\alpha\beta}$ includes $\theta_\alpha\theta_\beta$, which is anti-Hermitian as ($\theta_\alpha\theta_\beta)^\dagger=(\theta_\beta^\ast\theta_\alpha^\ast)^T=(\theta_\beta\theta_\alpha)^T=\theta_\alpha\theta_\beta=-\theta_\beta\theta_\alpha$. We note that the transpose defined by replacing subscripts of the elements of a matrix implies a property $(EF)^T=(-)^{\eps_F \eps_E}F^TE^T$.}
${\cal B}_{i\beta} $ and ${\cal C}_{\alpha j} $ are Grassmann-odd and related
as ${\cal C}^T=-{\cal B}$.  In the present paper, we assume that the bosonic submatrix of the kinetic matrix $A_{ij}$ is non-degenerate,
i.e., invertible. This assumption is equivalent to requiring
\begin{align}
 \det A_{ij}^{(0)} \neq 0 \ , \qquad {\rm where} \quad A_{ij}^{(0)} =A_{ij}|_{\tn, \tnd=0} \ ,
\end{align}
as discussed in Sec.~II.~A.

Multiplied by the inverse of $A_{ij}$, $A^{ij}$, the first line of (\ref{infmom}) can be solved for $\delta \qd^i$ as
\ba
\delta \qd^i= A^{ij} \delta p_j -A^{ij} \B_{j\alpha}\delta\tnd^\alpha
            - A^{ij} L_{\dot{q}^j q^k} \delta q^k + A^{ij} L_{\dot{q}^j \theta^\alpha} \delta \theta^\alpha 
 \ ,
\label{useful5}
\ea
and plugging this into the second line of (\ref{infmom}) gives 
\ba
(D_{\alpha\beta}-{\cal C}_{\alpha i}A^{ij}{\cal B}_{j\beta})
\delta\tnd^\beta = \delta \pi_\alpha - \C_{\alpha i} A^{ij} \delta p_j 
+\left( {\cal C}_{\alpha i}A^{ij} L_{\dot{q}^j q^k} - L_{\dot{\theta}^\alpha q^k} \right) \delta q^k 
-\left( {\cal C}_{\alpha i}A^{ij} L_{\dot{q}^j \theta^\beta} +
L_{\dot{\theta}^\alpha \theta^\beta} \right) \delta \theta^\beta
\ .
\label{tndEQN}
\ea
Now we would like to consider the situation such that the velocities
$\tnd^\alpha$ cannot be expressed in terms of other canonical variables,
that is, the coefficient matrix of $\delta\tnd^\alpha$ does not
have the inverse, equivalent to imposing the degeneracy condition,
\begin{align}
 \det D_{\alpha\beta}^{(0)}=0 \ , \qquad {\rm where} \quad D_{\alpha\beta}^{(0)} =D_{\alpha\beta}|_{\tn, \tnd=0} \ .
\end{align}
We consider the cases of $N = 1$, $N = 2$, and $N \ge 3$ separately.
\begin{itemize}

\item{\bf $N = 1$ case}: 

Let us start with a single fermionic variable, $N=1$.  In this case,
both $D$ and $\C A^{-1} \B$ are zero due to the Grassmann
property, and the degeneracy condition of the kinetic matrix is automatically
satisfied. More importantly, the coefficient matrix $D-\C A^{-1}\B$
always vanishes, and we have a primary constraint $\phi_1=\pi_1-f_1(q,p,\theta)$, which will remove the fermionic ghost properly.

\item{\bf $N = 2$ case}: 

When $N=2$, the matrix $D$ is no longer zero, which, in general, has the following form,
\begin{align}
D=
\left(
\begin{array}{cc}
	0 & D_{12} \\
	-D_{12} & 0 \\
\end{array}
\right) \ .
\end{align}
Let us explicitly write the Lagrangian for this case.
\begin{align}
 L=G_I(q^i,\dot{q}^i)x_I, \qquad 
 {\rm where} \qquad \bm{x}=
 \begin{pmatrix}
  1\\
  i\theta_1\theta_2\\
  i\theta_1\dot{\theta}_1\\
  i\theta_2\dot{\theta}_2\\
  i\theta_1\dot{\theta}_2\\  
  i\theta_2\dot{\theta}_1\\
  i\dot{\theta}_1\dot{\theta}_2\\
  \theta_1\theta_2\dot{\theta}_1\dot{\theta}_2 \\
 \end{pmatrix}\ ,
\end{align}
and $G_I$ $(I=1,2,\cdots,8)$ are functions depending on $q^i$ and $\dot{q}^i$ only. 
Therefore, we obtain
\begin{align}
 D_{12}=iG_7+G_8\theta_1\theta_2 \ .
\end{align}
Applying the degeneracy condition, we have $G_7=0$.
The momenta are now\footnote{
When we do not require the degeneracy condition, 
$iG_7\tnd_2$ and $-iG_7\tnd_1$, do appear in \eqref{momenta_N2}, which makes them solvable for $\tnd_2$ and $\tnd_1$, and extra degrees of freedom, corresponding to the fermionic ghost, remain in the fermionic sector.}
\begin{align}
 \pi_1=G_I\frac{\partial x_I}{\partial \dot{\theta}_1}=-iG_3\theta_1-iG_6\theta_2+G_8\theta_1\theta_2\dot{\theta}_2 \ ,\qquad
 \pi_2=G_I\frac{\partial x_I}{\partial \dot{\theta}_2}=-iG_5\theta_1-iG_4\theta_2-G_8\theta_1\theta_2\dot{\theta}_1 \ .
\label{momenta_N2}
\end{align}
The explicit form of $A_{ij}$ is 
\begin{align}
 A_{ij} = A_{ij}^{(0)}+\sum_{I>1}A_{ij}^I x_{I} \ , \qquad {\rm where}\qquad 
 A_{ij}^I=\frac{\partial^2 G_I}{\partial\dot{q}^j\partial\dot{q}^i}\ ,\quad A_{ij}^{(0)}=A_{ij}^1 \ .
\end{align}
The inverse is easily obtained since we have assumed $A_{ij}^{(0)}$ has the inverse:
\begin{align}
A^{jk}=&\left(\delta^j_l-A^{jm(0)}\sum_{I>1}A^I_{ml}x_I 
+A^{jm(0)}\sum_{I>1}A^I_{mn}x_IA^{nr(0)}\sum_{J>1}A^J_{rl}x_J\right)A^{lk(0)} \ .
\end{align}
Let us note that $x_I$ $(I>1)$ have at least one of $\theta_1$ and $\theta_2$ except for $I=7$, which does not contribute because of the degeneracy condition. Each component of the coefficient in the left-hand side of Eq.~\eqref{tndEQN}  is calculated straightforwardly and we have 
\begin{align}
 &D_{11}-\C_{1i}A^{ij}\B_{j1}=D_{22}-\C_{2i}A^{ij}\B_{j2}=0 \ , \nonumber\\
 &D_{12}-\C_{1i}A^{ij}\B_{j2}=\left[G_8+\left(-G_{3,\dot{q}^j}G_{4,\dot{q}^k}+G_{6,\dot{q}^j}G_{5,\dot{q}^k}\right)A^{jk(0)}\right]\theta_1\theta_2 \ ,\nonumber\\
 & D_{21}-\C_{2i}A^{ij}\B_{j1}=-(D_{12}-\C_{1i}A^{ij}\B_{j2}) \ .
\end{align}
Since $D-\C A^{-1}\B$ has terms with $\theta_1\theta_2$, we need to multiply \eqref{tndEQN} by $\theta_1$ or by $\theta_2$ to have relations among the canonical variables. For instance, we multiply it by $\theta_1$; however, we cannot have reasonable ones since they have $\theta_1 \delta \pi_\alpha$, which means  $\partial \pi_\alpha/ \partial z$ $(z=q^i,p^i,\theta_1,\theta_2)$ cannot be determined uniquely as we can add arbitrary functions proportional to $\theta_1$,
\begin{align}
 \frac{\partial \pi_\alpha}{\partial z}\rightarrow \frac{\partial \pi_\alpha}{\partial z}+g_{\alpha z}(q,p)\theta_1 \ .
\end{align}
Therefore, no phase space variable is properly constrained by these relations\,\footnote{This requirement is referred to as ``regularity condition'', where the Jacobian matrix of the $M'$ (independent) constraints with respect to the canonical variables should have rank $M'$, and hence, the constraints properly reduce the dimension of the phase space, as explained in \cite{Henneaux:1992ig}.}.
A quite similar discussion applies when we multiply \eqref{tndEQN} by $\theta_2$.
To avoid such a situation, the coefficient matrix of $\delta \dot{\theta}$ in (\ref{tndEQN}), $D - \C A^{-1} \B$, must vanish for the $N=2$ case, and we then have two primary constraints, 
\ba
\delta \phi_\alpha =\delta \pi_\alpha - \C_{\alpha i} A^{ij} \delta p_j 
+\Bigl( {\cal C}_{\alpha i}A^{ij} L_{\dot{q}^j q^k} - L_{\dot{\theta}^\alpha q^k} \Bigr) \delta q^k 
-\Bigl( {\cal C}_{\alpha i}A^{ij} L_{\dot{q}^j \theta^\beta} +
L_{\dot{\theta}^\alpha \theta^\beta} \Bigr) \delta \theta^\beta=0 \ ,
\ea
whose number is sufficient to eliminate half degrees of freedom in phase space.\\

\item{\bf $N \ge 3$ case}: 

Let us consider the $N \ge 3$ case.  In this case, the degenerate condition is no longer enough to eliminate all the extra degrees of freedom, and the analysis becomes quite involved. Thus, we just comment on the general analysis in Appendix~\ref{AppendixB} and concentrate on the case where all the extra degrees of freedom are eliminated only by primary constraints from now on, as similarly done in the bosonic case~\cite{Motohashi:2016ftl}.
Here we suppose that all the elements in the coefficient matrix of the left hand side in \eqref{tndEQN} vanish,
\ba 
D_{\alpha\beta}-{\cal C}_{\alpha i}A^{ij}{\cal B}_{j\beta}=0 \ ,
\label{DC1} 
\ea 
which yields $N$ primary constraints,
\ba
\delta \pi_\alpha - \C_{\alpha i}
     A^{ij} \delta p_j 
+ \Bigr( {\cal C}_{\alpha i}A^{ij} L_{\dot{q}^j q^k} - L_{\dot{\theta}^\alpha q^k} \Bigr) \delta q^k 
-\Bigr( {\cal C}_{\alpha i}A^{ij} L_{\dot{q}^j \theta^\beta} +
L_{\dot{\theta}^\alpha \theta^\beta} \Bigr) \delta \theta^\beta
= 0 \ .
\label{primaryd}
\ea
A straightforward calculation shows that they actually satisfy the integrable condition (including the case of $N=2$), and therefore, they have the integrated form,
\begin{align}
 \phi_\alpha=\pi_\alpha-F_\alpha(q,p,\theta)=0 \ .
\label{primary}
\end{align}
In Appendix~\ref{Appendix2}, we give an alternative proof of the equivalence of \eqref{DC1} and the existence of the primary constraints.
\end{itemize}
To summarize, the degeneracy condition, $\det D_{\alpha\beta}^{(0)}=0$, is equivalent to the maximally-degenerate condition, $D - \C A^{-1} \B = 0$, for the $N=1$ and $N=2$ cases. For $N \ge 3$, the latter one is a sufficient (but not necessary) condition for the former. In the following, we simply adopt the condition $D - \C A^{-1} \B = 0$ for any $N$.

\subsection{Total Hamiltonian and Dirac bracket}
\label{subsecHD}

We have obtained the condition generating $N$ primary constraints, which would
eliminate the fermionic  ghosts, for the Lagrangian with up to first time derivatives of $N$ fermions and $n$ bosons.  
In this subsection, we perform the explicit Hamiltonian analysis to find supplementary conditions for avoiding negative norm states.

Taking into account the primary constraints~\eqref{primary}, obtained from the maximally-degenerate condition~\eqref{DC1},
the total Hamiltonian
is given by
\ba
H_T=H+ \phi_\alpha \mu^\alpha \ ,
\ea
where the Hamiltonian $H$ is defined in (\ref{Hamiltonian}), and $\mu^\alpha$ are the Lagrange multipliers.
The variations of the Lagrangian including the constraints with respect to the canonical variables yield
\ba
\qd^i &=& \frac{\p H}{\p p_i} +  \frac{\p \phi_\alpha}{\p p_i}
\mu^\alpha \,\approx \{q^i,H_T\} \ , 
\qquad {\dot p}_i = -\frac{\p H}{\p q^i}-  \frac{\p \phi_\alpha}{\p
q_i} \mu^\alpha \,\approx \{p_i,H_T\} \ , \\
\qquad 
\tnd^\alpha &=&-\frac{\p H}{\p \pi_\alpha}-  \frac{\p \phi_\beta}{\p
\pi_\alpha} \mu^\beta \,\approx \{\theta^\alpha,H_T\} \ , 
\qquad {\dot \pi}_\alpha = -\frac{\p H}{\p \tn^\alpha}-  \frac{\p
\phi_\beta}{\p \tn^\alpha} \mu^\beta \,\approx \{\pi_\alpha,H_T\} \ .
\label{HamiltonEq2}
\ea
The time derivative of the primary constraints~(\ref{primary}) is given by
\ba
\dot{\phi}_\alpha \approx \{\phi_\alpha, H_T\} \approx
\{\phi_\alpha, H\} + \{\phi_\alpha, \phi_\beta\}\mu^\beta
\approx 0 \ ,
\ea
where we have used the identity~(\ref{P2}). 
Since we have a sufficient number of constraints, we assume that 
the Poisson brackets between the primary constraints,
\ba
C_{\alpha\beta} \equiv \{\phi_\alpha,\phi_\beta\}= \frac{\p F_\alpha}{\p q^i} \frac{\p F_\beta}{\p p_i}- \frac{\p F_\alpha}{\p p_i} \frac{\p F_\beta}{\p q^i}+ \frac{\p F_\alpha}{\p \tn^\beta}+ \frac{\p F_\beta}{\p \tn^\alpha} \ ,
\label{PC}
\ea
have their inverse, 
\begin{align}
 \det C_{\alpha\beta}^{(0)}\neq 0 \ , 
\label{second_hamilton}
\end{align} 
where
\begin{align}
  C_{\alpha\beta}^{(0)}=\left.C_{\alpha\beta}\right|_{\theta=0}=\Bigl(\frac{\partial F_\alpha}{\partial \theta^\beta}\Bigr)^{(0)}+\Bigl(\frac{\partial F_\beta}{\partial \theta^\alpha}\Bigr)^{(0)}\ ,
\end{align}
which means all the primary constraints~(\ref{primary}) are second class\,\footnote{It should be noticed that, even if we distinguish the maximal number of first class constraints, the number of second class constraints for fermionic system is not necessarily even, in sharp contrast with bosonic case. This is because the Poisson brackets between fermionic variables are not anti-symmetric but symmetric on the replacement of the variables.}\,\footnote{Let us mention the case where the Poisson brackets are not invertible, i.e., $\det \{\phi_\alpha,\phi_\beta\}^{(0)}=0$. In the usual Dirac's algorithm, we immediately have (at most) $N-r$ secondary constraints, where $r$ is the rank of $\{\phi_\alpha,\phi_\beta\}$. In our case, where we, in general, have fermionic non-linear terms in the Lagrangian, we need several conditions to have secondary constraints and definite dynamics. Please see Appendix~\ref{Appendix3} for further discussion.}. As a result, all the Lagrange multipliers $\mu^\alpha$ are fixed, and no further constraints appear.
The total number of degrees of freedom is now $(2(n+N)-N)/2=n+N/2$ as desired.
The Dirac brackets between $\tn^\alpha$ are given by
\ba
\{\tn^\alpha, \tn^\beta\}_D=\{\tn^\alpha, \tn^\beta\}-\{\tn^\alpha,\phi_\gamma\}(C^{-1})^{\gamma\delta}\{\phi_\delta,\tn^\beta\}=-(C^{-1})^{\alpha\beta} \ .
\label{DB}
\ea
By virtue of the Dirac brackets, other relations including $\pi$ are expressed only in terms of $q$, $p$ and $\theta$ as
\begin{align}
 \{\theta^\alpha, \pi_\beta\}_D=\{\theta^\alpha, F_\beta\}_D \ , \qquad 
 \{\pi_\alpha, \pi_\beta\}_D=\{F_\alpha, F_\beta\}_D \ ,
\label{dirac_pi_F}
\end{align}
which implies that the degrees of freedom corresponding to $\pi$ are completely eliminated from the dynamics and we need not consider them after the quantization.
(All the Dirac brackets between the canonical variables are calculated in Appendix~\ref{Appendix4}.)
Following the quantization prescription, we have
\begin{align}
 \{\tnh_\alpha, \tnh_\beta\}_+ = -i(C^{-1}(\hat{q}, \hat{p}, \tnh))_{\alpha\beta} \ .
\end{align}
Though, in general, the matrix $C_{\alpha\beta}$ is a function of all the canonical variables, let us assume that it depends only on bosonic variables $(q, p)$ for a concrete statement. Then, as long as all the eigenvalues of $-i C^{-1}_{\alpha \beta}$, or of $i C_{\alpha\beta}$, are positive definite, any fermionic states have their positive norm. As a consequence, we have obtained a set of sufficient conditions for avoiding negative norm states,
which are, more concretely, the maximally-degenerate condition~$D - \C A^{-1} \B = 0$, $\det\{\phi_\alpha,\phi_\beta\}^{(0)}\neq 0$, and positive definiteness  of $i \{\phi_\alpha, \phi_\beta\}$ (with suitable initial values of canonical variables solving the constraints). 
\subsection{Sufficient conditions in Lagrangian formulation}

In this subsection, we derive the relations between the obtained conditions in Hamiltonian formulation and the equations of motion derived in Lagrangian formulation. It becomes clear that the maximally-degenerate condition for the absence of the fermionic ghosts guarantees that the equations of motion for fermions are first-order differential equations. We also express the Poisson brackets between the constraints in terms of the equations of motion and find that the invertibility of the Poisson brackets, which is responsible for keeping the number of degrees of freedom $n+N/2$, is equivalent to the condition that the equations of motion for fermions be uniquely solved for the first derivatives of fermions.

The Euler-Lagrange equations~(\ref{ELeqn}), derived from the Lagrangian~(\ref{Lagrangian}), can be 
written as
\ba
K
\left(
\begin{array}{c}
	\qdd^j  \\
	\tndd^\beta  \\
\end{array}
\right)
=\left(
\begin{array}{c}
	E_i  \\
	{\cal E}_\alpha  \\
\end{array}
\right) \ ,
\label{EOM}
\ea
where we have defined a Grassmann-even column vector $E_i$ and a Grassmann-odd one ${\cal E}_\alpha$ as
\begin{align}
 E_i (q^i, \qd^i, \tn^\alpha, \tnd^\alpha)&= L_{q^i}-\dot{q}^j L_{\dot{q}^i q^j}-\dot{\theta^\alpha}L_{\dot{q}^i \theta^\alpha} \ , \label{eq:Ei}\\
{\cal E}_\alpha (q^i, \qd^i, \tn^\alpha, \tnd^\alpha)&=L_{\theta^\alpha}-\dot{q}^i L_{\dot{\theta}^\alpha q^i}-\dot{\theta}^\beta L_{\dot{\theta}^\alpha\theta^\beta}\ . \label{eq:Ealpha}
\end{align}
Then, the invertibility of the matrix $A_{ij}$ suggests 
that the first line of (\ref{EOM}) can be rewritten as
\ba
\qdd^i = A^{ij}E_j -A^{ij} \B_{j\beta}\tndd^\beta \ .
\label{EOMqdd}
\ea
Plugging (\ref{EOMqdd}) into the second line of (\ref{EOM}), 
we obtain second-order differential equations for fermions,
\ba
(D_{\alpha\beta}-{\cal C}_{\alpha i}A^{ij}{\cal B}_{j\beta}) \tndd^\beta = \E_\alpha- \C_{\alpha i} A^{ij} E_j \ .
\label{EOMtheta}
\ea
When we impose the maximally-degenerate condition~(\ref{DC1}), the left-hand side of (\ref{EOMtheta}) vanishes, and the equations of motion for fermions become first-order differential equations, 
\ba
\Y_\alpha (q^i, \qd^i, \tn^\alpha, \tnd^\alpha) \equiv  {\cal E}_\alpha - {\cal C}_{\alpha i } A^{ij} E_j = 0 \ .
\label{EOMf}
\ea
Here, the first-order equations should be solved for $\tnd^\alpha$ in order for $\dot{\theta}^\alpha$ to be uniquely determined. Then, the inverse function theorem suggests another condition,
which we call the uniqueness condition,
\ba\
\det  J^{(0)}_{\alpha\beta} \neq 0 \ , 
\label{DC2}
\ea
where we defined a Grassmann-even matrix,
\begin{align}
 J_{\alpha\beta}=\left.\frac{\partial  \Y_\alpha}{ \partial \tnd^\beta} \right|_{q,\qd,\tn}\ .
\end{align}

To see that the equations of motion for $q^i$ remain second-order differential equations, 
we first take the time derivative of (\ref{EOMf}), 
\ba
{\dot  \Y_\alpha } =\qdd^i \, { \partial \Y_\alpha \over \partial \qd^i}+ \tndd^\beta { \partial \Y_\alpha \over \partial \tnd^\beta}+  \qd^i {\partial  \Y_\alpha \over \partial q^i}+ \tnd^\beta{\partial  \Y_\alpha \over \partial \tn^\beta}=0 \ .
\label{EOMYd}
\ea
Since we have imposed (\ref{DC2}), one can solve this expression for $\tndd^\alpha$.
Then, substituting this into the equations (\ref{EOMqdd}), we obtain 
the second-order differential equations for $q^i$,
\ba
\qdd^k \left(\delta^i_{~k}-A^{ij}\B_{j\beta} J^{\beta\gamma}  { \partial \Y_\gamma \over \partial \qd^k}\right)=
A^{ij}E_j+A^{ij}\B_{j\beta}J^{\beta\gamma}\left(  \qd^k {\partial  \Y_\gamma \over \partial q^k}+ \tnd^\delta{\partial  \Y_\gamma \over \partial \tn^\delta}\right).
\ea
One can immediately notice that the time evolution of $q^i$ is uniquely determined 
since the coefficient matrix of $\qdd^i$ is invertible. 
Therefore, we find that the bosons and the fermions respectively obey
second-order and first-order equations,
\ba
 \qdd^i=W^i(q^j, \qd^j, \tn^\beta) \ , \qquad   \tnd^\alpha=Z^\alpha(q^j, \qd^j, \tn^\beta) \ ,
\ea
where $W^i$ and $Z^\alpha$ are even and odd functions of $q^i, \qd^i$, and $\tn^\alpha$.
Thus, the number of the initial conditions needed to solve these equations is $2n+N$, which agrees with the dimension of the phase space analyzed in Hamiltonian formulation.

Now we would like to show that the above condition~\eqref{DC2} is equivalent to the invertibility of the Poisson brackets between the primary constraints, introduced as \eqref{second_hamilton}. 
As in the case of bosons~\cite{Motohashi:2016ftl}, 
we will make use of the primary constraints~(\ref{primary}). 
Recall that $\pi_\alpha= \p L / \p \tnd^\alpha$ and $p_i=\p L / \p \qd^i$.
The derivatives of the constraints, $\pi_\alpha = F(q^j, p_j, \tn_\beta)$, with respect to $q^j, \qd^j, \tn^\beta,$ and $\tnd^\alpha$ can be written as
\ba
L_{\tnd^\alpha\tnd^\beta} &=& L_{\qd^i \tnd^\beta} {\p F_\alpha \over \p p_i} \ , \label{useful1}\\
L_{\tnd^\alpha\tn^\beta} &=& L_{\qd^i \tn^\beta} {\p F_\alpha \over \p p_i}+ {\p F_\alpha \over \p \tn^\beta} \ , \label{useful2}\\
L_{\tnd^\alpha \qd^j} &=& L_{\qd^i \qd^j} {\p F_\alpha \over \p p_i} \ , \label{useful3}\\
L_{\tnd^\alpha q^j} &=& L_{\qd^i q^j} {\p F_\alpha \over \p p_i} + {\p F_\alpha \over \p q^j} \ .\label{useful4}
\ea
Plugging these relations into (\ref{EOMf}) through (\ref{eq:Ei}) and (\ref{eq:Ealpha}), 
we obtain the explicit expression for $\Y_\alpha$,
\ba
\Y_\alpha = L_{\tn^\alpha} -\qd^i  {\p F_\alpha \over \p q^i} -\tnd^\beta {\p F_\alpha \over \p \tn^\beta} - {\p F_\alpha \over \p p_i}  L_{q^i} \ .
\ea
Let us calculate 
\begin{align}
 \left.\frac{\partial \Y_\alpha}{\partial \dot{\theta}^\beta}\right|_{q, p, \theta}
 =  \left.\frac{\partial \Y_\alpha}{\partial \dot{\theta}^\beta}\right|_{q, \dot{q}, \theta}
 +\left.\frac{\partial \dot{q}^i}{\partial \dot{\theta}^\beta}\right|_{q,p,\theta}
 \left.\frac{\partial \Y_\alpha}{\partial \dot{q}^i}\right|_{q,\theta,\dot{\theta}} \ .
\label{derivative_p_q}
\end{align}
Here we note that 
\begin{align}
 \left.\frac{\partial \dot{q}^i}{\partial \dot{\theta}^\beta}\right|_{q,p,\theta}
 = A^{ij} \B_{j\beta}=-\frac{\partial F_\beta}{\partial p_i} \ ,
\label{useful6}
\end{align}
where we have used \eqref{useful5} and \eqref{useful3}.
The left hand side of \eqref{derivative_p_q} is explicitly calculated as 
\begin{align}
 \left.\frac{\partial \Y_\alpha}{\partial \dot{\theta}^\beta}\right|_{q, p, \theta}
 &=\Bigl(L_{\theta^\alpha\dot{\theta}^\beta}+\frac{\partial \dot{q}^i}{\partial \dot{\theta}^\beta}L_{\theta^\alpha\dot{q}^i}\Bigr)-\frac{\partial \dot{q}^i}{\partial \dot{\theta}^\beta}\frac{\partial F_\alpha}{\partial q^i}-\frac{\partial F_\alpha}{\partial\theta^\beta}+\frac{\partial F_\alpha}{\partial p^i}\Bigl(L_{q^i\dot{\theta}^\beta}+\frac{\partial \dot{q}^j}{\partial \dot{\theta}^\beta}L_{q^i\dot{q}^j}\Bigr)\nonumber\\
 &=\Bigl(-L_{\qd^i \tn^\alpha} {\p F_\beta \over \p p_i}-{\p F_\beta \over \p \tn^\alpha}+\frac{\partial \dot{q}^i}{\partial \dot{\theta}^\beta}L_{\theta^\alpha\dot{q}^i}\Bigr)-\frac{\partial \dot{q}^i}{\partial \dot{\theta}^\beta}\frac{\partial F_\alpha}{\partial q^i}-\frac{\partial F_\alpha}{\partial\theta^\beta}+\frac{\partial F_\alpha}{\partial p^i}\Bigl(L_{\qd^j q^i} {\p F_\beta \over \p p_j} + {\p F_\beta \over \p q^i}+\frac{\partial \dot{q}^j}{\partial \dot{\theta}^\beta}L_{q^i\dot{q}^j}\Bigr)\nonumber\\
 &=-\left(\frac{\p F_\alpha}{\p q^i} \frac{\p F_\beta}{\p p_i}-
\frac{\p F_\alpha}{\p p_i} \frac{\p F_\beta}{\p q^i}+ \frac{\p
	F_\alpha}{\p \tn^\beta}+ \frac{\p F_\beta}{\p \tn^\alpha}\right)
 =-\{\phi_\alpha,\phi_\beta\}  \ ,
\end{align}
where we have used \eqref{useful2} and \eqref{useful4} in the second line, and \eqref{useful6} in the third line.
Therefore, we find 
\begin{align}
 \left.\frac{\partial \Y_\alpha}{\partial \dot{\theta}^\beta}\right|_{q, p, \theta}
 =-\{\phi_\alpha,\phi_\beta\} 
 =\left.\frac{\partial \Y_\alpha}{\partial \dot{\theta}^\beta}\right|_{q, \dot{q}, \theta}
 +A^{ij}\B_{j\beta}
 \left.\frac{\partial \Y_\alpha}{\partial \dot{q}^i}\right|_{q,\theta,\dot{\theta}}  \ ,
\end{align}
where we again used \eqref{useful6} in the right hand side of \eqref{derivative_p_q}, and 
\begin{align}
 -C_{\alpha\beta}^{(0)}= J_{\alpha\beta}^{(0)} \ .
\end{align}
As a result, we explicitly see that the invertibility of $C_{\alpha\beta}$, \eqref{second_hamilton}, coincides with the non-zero determinant of $J_{\alpha\beta}^{(0)}$, the uniqueness condition~\eqref{DC2}.
In other words, the condition that all the Lagrange multipliers be uniquely fixed 
is equivalent to the condition that the time evolution of the system be uniquely determined.

\section{Concrete models}
\label{section5}

In the previous section, we have derived the conditions to successfully eliminate
unwanted degrees of freedom in the fermionic sector.  
In this section, we provide some examples of a degenerate (boson-)fermion system, having $n+N/2$ physical degrees of freedom.\\

{\bf \textrm{Example 1 :}}
Let us first consider the simplest example, where the bosonic sector is absent. 
In this case, one can immediately notice that the Lagrangian should be linear in the time derivatives of fermions from the maximally-degenerate condition~\eqref{DC1}. Then, the most general Lagrangian in this case is given by
\ba
L=i f_\alpha(\tn^\beta) \tnd^\alpha \ ,
\ea
where $f_\alpha$ are arbitrary Grassmann-odd functions of $\tn^\beta$.
The momenta are easily found as 
\begin{align}
 \pi_\alpha=-if_\alpha \ ,
\end{align}
which lead to the constraints, $\phi_\alpha = \pi_\alpha + i f_\alpha$.
As long as the matrix,
\ba
C_{\alpha\beta} = -i\frac{\p f_\alpha}{\p \tn^\beta}-i \frac{\p f_\beta}{\p \tn^\alpha} \ ,
\ea
is invertible, the number of degrees of freedom is $N/2$. 

~\\
{\bf \textrm{Example 2 :}} 
The second example is a Lagrangian for $n=1$ and $N=2$,
\ba
L={1 \over 2} \qd^2 + i \qd (\tn_1+\tn_2) \tnd_1\ ,
\ea
which satisfies the condition~(\ref{DC1}). The momenta are given by
\ba
p= \qd + i(\tn_1 + \tn_2) \tnd_1 \ , \quad
\pi_1= -i\qd(\tn_1 + \tn_2) \ ,\quad
\pi_2=0 \ ,
\ea
where the last two lead to the primary constraints, $\phi_1 = \pi_1 + ip(\theta_1+\theta_2)$ and $\phi_2 =\pi_2$. Then, the constraint matrix $C_{\alpha\beta}$ is invertible
(for a non-zero value of $p$)
since
\ba
\det C_{\alpha\beta} = p^2 \ .
\ea
Thus, the total number of degrees of freedom is $2=1+2 \times 1/2$.

~\\
{\bf \textrm{Example 3 :}}
An example for $n=1$ and arbitrary $N$ is given by
\ba
L={1 \over 2} \qd^2 
+ i\bigl(f_1(q, \theta^\beta)+f_2(q, \theta^\beta)\qd\bigr) \tn_\alpha \tnd^\alpha 
+{1\over 2}g(q,\theta^\gamma)
\tn_\alpha \tn_\beta \tnd^\alpha \tnd^\beta \ .
\ea
The maximally-degenerate condition,
\begin{align}
 L_{\dot{\theta}^\alpha\dot{\theta}^\beta}+L_{\dot{\theta}^\alpha\dot{q}}L^{-1}_{\dot{q}\dot{q}}L_{\dot{q}\dot{\theta}^\beta}=\bigl(g-(f_2)^2\bigr)\theta_\alpha\theta_\beta=0 \ ,
\end{align}
implies $g=f_2^2$, which we suppose from now on. The conjugate momenta are
\begin{align}
 p=\dot{q}+if_2\theta_\alpha\dot{\theta}^\alpha \ ,
\end{align}
\begin{align}
 \pi_\alpha=-i(f_1+f_2\dot{q})\theta_\alpha+g \theta_\alpha\theta_\beta\dot{\theta}^\beta
 =-i(f_1+f_2 p) \theta_\alpha \ ,
\end{align}
where the last line is again regarded as the primary constraints, $\phi_\alpha =
\pi_\alpha+i(f_1+f_2 p)\theta_\alpha$. As long as $\left.(f_1+f_2 p)\right|_{\theta=0}=f^{(0)}_1+f^{(0)}_2 p\neq 0$, the constraint matrix $C_{\alpha\beta}$ is invertible since
\ba
\det C_{\alpha\beta}^{(0)} = \bigl(-2i(f_1^{(0)}+f_2^{(0)} p)\bigr)^N \ .
\ea
Then, the system has $N$ second class constraints, and the total number of degrees of freedom is $1+N/2$. 

~\\
{\bf \textrm{Example 4 :}}
A similar but practically different model to the previous one is
\begin{align}
 L=\frac{1}{2}(\dot{q}+i\epsilon_{\alpha\beta}\theta^\alpha\dot{\theta}^\beta)^2+\frac{i}{2}\theta_\alpha\dot{\theta}^\alpha \ .
\end{align}
We should note that it is not an essentially new model since there exists an invertible transformation as $q\rightarrow q+(i/2)\epsilon_{\alpha\beta}\theta^\alpha{\theta}^\beta$ and $\theta^\alpha\rightarrow\theta^\alpha$ between this Lagrangian and $L=\qd^2/2 + (i/2) \,\tn_\alpha \tnd^\alpha$. However, it would be worthwhile to examine this model because we have found a field theoretical extension of this model as exhibited in Appendix~\ref{Appendix5}.
The conjugate momenta are 
\begin{align}
 p=\dot{q}+i\epsilon_{\alpha\beta}\theta^\alpha\dot{\theta}^\beta \ ,
\end{align}
\begin{align}
 \pi_{\alpha}=i\epsilon_{\alpha\beta}\theta^\beta(\dot{q}+i\epsilon_{\gamma\delta}\theta^\gamma\dot{\theta}^\delta)-\frac{i}{2}\theta_\alpha=i\epsilon_{\alpha\beta}\theta^\beta p - \frac{i}{2}\theta_\alpha \ .
\end{align}
Therefore, $N$~primary constraints are found as $\phi_\alpha=\pi_\alpha-i\epsilon_{\alpha\beta}\theta^\beta p +(i/2)\theta_\alpha$. The Poisson brackets,
\begin{align}
 \{\phi_\alpha,\phi_\beta\}=-i\delta_{\alpha\beta} \ ,
\end{align}
imply the invertibility since 
\begin{align}
 \det C_{\alpha\beta}=(-i)^N \ .
\end{align}
As a result, the number of degrees of freedom is $1+N/2$.
	If the canonical kinetic term, $(i/2)\theta_\alpha\dot{\theta}^\alpha$, is absent, 
	this system will generate secondary constraints and/or have first class constraints 
	since $\{\phi_\alpha, \phi_\beta\}$ vanishes. The use of $(i/2)\epsilon_{\alpha\beta}\theta^\alpha\dot{\theta}^\beta$ instead of $(i/2)\theta_\alpha\dot{\theta}^\alpha$ also gives the vanishing Poisson brackets and does not work as well. In those cases, we will have a smaller number of degrees of freedom than $1+N/2$, which shows the explicit difference from Example~3. 
In the field theoretical extension given in Appendix \ref{Appendix5}, 
the standard Weyl kinetic term plays the same role with $(i/2)\theta_\alpha\dot{\theta}^\alpha$.

\vspace{1mm}

\section{Summary}

As mentioned in \cite{Henneaux:1992ig}, even when the Lagrangian contains only up to first derivatives of fermions, a
non-degenerate fermionic system always suffers from the presence of negative norm states,
which come as a consequence of the existence of extra degrees of freedom.
Although the situation in the fermionic case is more involved because of the Grassmann property of fermionic variables,
this can be contrasted with a non-degenerate bosonic system containing second or higher derivatives
in the Lagrangian.
In such bosonic system, the Hamiltonian should include terms linear in momentum, 
making the Hamiltonian unbounded from below. This is what is called Ostrogradsky's ghost instability. 
So far, there seem to be no obvious criteria to determine the existence of the ghosts in fermionic system at the classical level, which are, in turn, transparently observed as negative norm states once the system is quantized. (The relation between Ostrogradsky's ghosts and negative norm states is more obvious in a bosonic system as shown in \cite{Woodard:2015zca}.)
To avoid these negative norm states, the fermionic system must be degenerate and contain a sufficient number of constraints to eliminate half degrees of freedom in phase space, 
whose situation is similar for a bosonic Lagrangian including second derivatives as investigated in \cite{Motohashi:2016ftl}.

In this paper, we have investigated extended fermionic theories non-trivially coupled with healthy bosons in the context of a point particle system. 
In Hamiltonian formulation, we have explicitly shown the maximally-degenerate condition to have $N$ primary constraints, which possibly lead to an appropriate number of degrees of freedom, $n+N/2$, and remove fermionic ghosts. The condition is that all the components of $D-\C A^{-1} \B$ vanish, which looks quite similar to that in a degenerate bosonic system. We have also obtained another condition, the Poisson brackets between the primary constraints must be invertible, to complete the Hamiltonian analysis.
This is not only because we need not have secondary constraints since we already have a sufficient number of constraints,
but also because the definite time evolution of the system is not guaranteed in a fermionic system.
It is noteworthy that such a doubt about whether we have the unique time evolution from a set of initial conditions without any ambiguity other than gauge degrees of freedom never appears in a purely bosonic system and is specific to the system including fermions. 
In Lagrangian formulation, we have derived equations of motion and found that satisfying the maximally-degenerate condition is equivalent to the condition that all the fermionic equations of motion be first-order differential equations. There, we have also shown that the invertibility of the Poisson brackets between the primary constraints coincides with the uniqueness condition that all the velocities of fermions be uniquely determined by the $N$ first-order differential equations, i.e., the time evolution is uniquely solved as mentioned above.
As a result, we conclude that, when both of the conditions are satisfied, primary constraints properly reduce the dimension of the phase space to $2n+N$, and correspondingly lead to $n+N/2$ physical degrees of freedom as desired. We have also provided some interesting examples, satisfying the conditions we derived in the general framework. As a special case where we have only fermionic variables, the Lagrangian should be linear in the time derivative of fermionic variables, which results in just a simple extension of Weyl fermions. Once fermionic variables are coupled to bosonic ones, their nonlinear derivative interaction comes in, and a variety of extensions, most of which might never have been considered, become possible, as some are explicitly given in the text.
Our analysis suggests the possibility of nonlinear extension to higher derivatives of fermions, and it is natural to next consider a further extension to Lorentz-invariant theories. 
As an implication to such field theoretical extension, we have presented a Lorentz-invariant ghost free boson-fermion co-existence Lagrangian in Appendix~\ref{Appendix5}.
The full analysis in the context of field theory will be reported soon in a future work~\cite{Kimura:2017}.

\acknowledgments 

We would like to thank Takahiro Tanaka and Daisuke Yoshida for useful discussions.
We would also like to thank Marc Heneaux and Richard Woodard for useful communications.
This work was supported in part by JSPS Grant-in-Aid for Scientific
Research Nos.~25287054 (R.K.~\&~M.Y.) and 26610062 (M.Y.), MEXT
Grant-in-Aid for Scientific Research on Innovative Areas ``New
developments in astrophysics through multimessenger observations of
gravitational wave sources'' No.~24103006.(Y.S.) and ``Cosmic
Acceleration'' No.~15H05888 (M.Y.).

\appendix

\section{Detailed analysis in partially-degenerate case}
\label{AppendixB}

Even if we do not impose the maximally-degenerate condition (\ref{DC1}), there is room to correctly remove the extra degrees of freedom and to have ghost-free action. One possibility is to have first class constraints, and the other is to generate secondary or further constraints. Of course, both of them might be realized simultaneously. 
In this appendix, let us consider such possibilities. Though we do not have practical examples for them here, we obtain necessary conditions for the Lagrangian when we apply such the partially-degenerate case.

Since $D_{\alpha\beta}^{(0)}$ is an anti-Hermitian matrix\footnote{One
should note that the degeneracy condition $\det D_{\alpha\beta}^{(0)}=0$
is automatically satisfied thanks to the Jacobi's theorem when the
number of fermions is odd.}, it can be diagonalized by a unitary matrix $P_{\alpha\beta}(q,p)$. Multiplying (\ref{tndEQN}) by $P^{-1}$,
we obtain
\ba
P^{-1} (D- \C A^{-1} \B) P  \, P^{-1} \delta\tnd 
= P^{-1} \left(\delta \pi - \C  A^{-1} \delta p
+ ( {\cal C}A^{-1} L_{\dot{q} q} - L_{\dot{\theta} q} ) \delta q 
-( {\cal C}A^{-1} L_{\dot{q} \theta} + L_{\dot{\theta} \theta} ) \delta \theta \right)
\ , 
\label{Diag_const0}
\ea
where we omitted the indices for simplicity.  
Then, the matrix in the left hand side has the following form,
\begin{align}
	P^{-1} (D- \C A^{-1} \B) P=
	\begin{pmatrix}
		R_{m\times m} & S_{m\times (N-m)}\\
		T_{(N-m)\times m} & U_{(N-m)\times (N-m)}
	\end{pmatrix} \ ,
	\label{Diag_const}
\end{align}
where 
\begin{align}
	R^{(0)}= {\rm diag}\{\lambda_1, ..., \lambda_m\}\ , \quad S^{(0)}=0 \ , \quad T^{(0)}=0\ , \quad U^{(0)}=0 \ .
\end{align}
We note that $m$ is the rank of $D_{\alpha\beta}^{(0)}$, and $\lambda_1,..., \lambda_m$ are non-zero eigenvalues of $D_{\alpha\beta}^{(0)}$.
We rewrite \eqref{Diag_const0} as
\begin{align}
	\begin{pmatrix}
		R_{IJ} & S_{I {\tt b}} \\
		T_{{\tt a} J} & U_{\tt ab}
	\end{pmatrix}
	\begin{pmatrix}
		(P^{-1}\delta \dot{\theta})_{J} \\
		(P^{-1}\delta \dot{\theta})_{\tt b}
	\end{pmatrix}
	=
	\begin{pmatrix}
		(P^{-1} \delta \tilde{\phi})_{I}\\
		(P^{-1} \delta \tilde{\phi})_{\tt a}
	\end{pmatrix} ,
	\label{Matrix_Diag}
\end{align}
where 
\begin{align}
	\delta \tilde{\phi}=\delta \pi - \C  A^{-1} \delta p
	+ ( {\cal C}A^{-1} L_{\dot{q} q} - L_{\dot{\theta} q} ) \delta q 
	-( {\cal C}A^{-1} L_{\dot{q} \theta} + L_{\dot{\theta} \theta} ) \delta \theta \ ,
\label{def_tilphi}
\end{align}
$I,J=1,\cdots,m$, and ${\tt a, b}=m+1,\cdots,N$. (We have omitted the subscripts in \eqref{def_tilphi}.)
Since $R$ is invertible, one can solve the first line for $(P^{-1}\delta \dot{\theta})_J$, and hence, this does not produce any constraints.
On the other hand, 
eliminating $(P^{-1}\delta \dot{\theta})_J$ in the second line of (\ref{Matrix_Diag}) by using the first line,
we find 
\begin{align}
	(U_{\tt ab}-T_{{\tt a}I}(R^{-1})^{IJ}S_{J {\tt b}})(P^{-1}\delta \dot{\theta})_{\tt b}
	=(P^{-1}\delta \tilde{\phi})_{\tt a}-T_{{\tt a}I}(R^{-1})^{IJ}(P^{-1}\delta \tilde{\phi})_{J} \ .
	\label{second_condition}
\end{align}
Note that $U-TR^{-1}S$ has no purely bosonic part and starts from quadratic terms of fermionic variables and their time derivatives.
As mentioned in the case of $N=2$,
we need to multiply \eqref{second_condition} by fermionic variables to remove the dependence on $\dot{\theta}$.
Then, these are not acceptable constraints 
because we have the freedom to add arbitrary functions to shift partial derivatives of $\pi$.\,\footnote{In general, nonlinear terms in $\dot{\theta}$ may appear in the left hand side of \eqref{second_condition}, which do not depend on any $\theta$. In these cases, we do not have even corresponding ``apparent constraints'' but some equations, which we can neither solve for $\delta \dot{\theta}$ nor constrain any canonical variables with. As a result, none of the equations in \eqref{second_condition} can be regarded as constraints. }
Therefore, 
we require
\begin{align}
	U_{\tt ab}-T_{{\tt a}I}(R^{-1})^{IJ}S_{J {\tt b}}=0 \ , 
	\label{Condition_A}
\end{align}
which generate up to $N-m$ primary constraints,
\begin{align}
	[(P^{-1})^{{\tt a}\alpha}-T_{{\tt a}I}(R^{-1})^{IJ}(P^{-1})^{J \alpha}]\delta \tilde{\phi}_\alpha=0 \ .
\end{align}
As a result, $(P^{-1}\delta \dot{\theta})_{\tt b}$ cannot be determined, 
and therefore, in order for $(P^{-1}\delta \dot{\theta})_J$ to be determined 
by the first line of \eqref{Matrix_Diag}, the dependence on $(P^{-1}\delta \dot{\theta})_{\tt b}$ 
should vanish, i.e., $S=0$. 
Then, $U=0$ follows from the condition~\eqref{Condition_A}. 
Since $P^{-1}(D-\C A^{-1}B)P$ is anti-Hermitian, $T=-S^{\dagger}$ holds, and $T=0$. 
Thus, the condition $S=T=U=0$ is required. Furthermore, to complete the Hamiltonian analysis, we need to check if there are secondary constraints. Please see Appendix~\ref{Appendix3} for the following procedure.

\section{Note on Dirac's algorithm of Hamiltonian mechanics with fermionic variables}
\label{Appendix3}
In this appendix, we would like to see how Dirac's algorithm is modified for a fermionic system from the usual one when the Poisson brackets between primary constraints are not invertible, $\det \{\phi_\alpha,\phi_\beta\}^{(0)}=0$.
As in purely bosonic cases,
after we obtain primary constraints $\phi_\alpha(\alpha=1,\cdots,m')$, we calculate the consistency conditions,
\begin{align}
 \dot{\phi}_\alpha=\{\phi_\alpha,
 H\}+\{\phi_\alpha,\phi_\beta\}\lambda^\beta ~\approx 0\,,
\end{align}
where $\lambda^\alpha$ are Lagrange multipliers. By taking the linear combinations of the constraints, we can redefine the constraints and the Lagrange multipliers as 
\begin{align}
 \dot{\phi}_\alpha=\{\phi_\alpha, H\}+
\begin{pmatrix}
 V_{r\times r} & W_{r\times (m'-r)}\\
 X_{(m'-r)\times r} & Y_{(m'-r)\times(m'-r)}
\end{pmatrix}_{\alpha\beta}
\lambda^\beta ~\approx 0\,, 
\end{align}
where 
\begin{align}
 \det V^{(0)}\neq 0 \ , \quad W^{(0)}=0 \ , \quad X^{(0)}=0 \ , \quad Y^{(0)}=0 \ ,
\end{align}
and $r$ is the rank of $\{\phi_\alpha,\phi_\beta\}^{(0)}$. The difference from the usual cases is that $W$, $X$ and $Y$ can have fermionic components. (Of course, $V$ also can have fermionic components as well, but it is not so important here.) We can discuss it in a similar manner to Appendix~\ref{AppendixB}. 
Let us decompose $\lambda^\alpha$ into $(\lambda, \bar{\lambda})^T$ and $\{\phi_\alpha, H\}$ into $(A,B)^T$, symbolically. Then, the first line multiplied by $V^{-1}$ is written as
\begin{align}
 \lambda ~\approx~ -V^{-1}W\bar{\lambda}-V^{-1}A \ . 
\label{det_multiplier_upper}
\end{align}
If we substitute this into the second line,
\begin{align}
 B+X\lambda+Y\bar{\lambda} ~\approx~ 0 \ ,
\end{align}
we have
\begin{align}
 (Y-X V^{-1}W)\bar{\lambda} ~\approx~ -B+X V^{-1}A \ .
\label{Apparent_constraint}
\end{align}
As before, $Y-X V^{-1} W$ does not have purely bosonic components, and
therefore, $\bar{\lambda}$ are not uniquely determined in any practical ways.\footnote{When, for some lines of \eqref{Apparent_constraint}, the right hand side is not weakly zero, $-B+X V^{-1}A\not\approx 0$, and the left hand side does not vanish, $Y-XV^{-1}W \ne 0$, they lead to (apparent) secondary constraints. However, these constraints violate the regularity condition, and then, we can rule out the possibility of having non-vanishing components in the corresponding lines of $Y-XV^{-1}W$. On the other hand, such discussion does not apply for the (weakly-)vanishing lines of $-B+X V^{-1}A$. For an extreme example, where all lines of $-B+X V^{-1}A$ completely vanish in the weak sense, the regularity condition is useless to rule out the possibility of having the non-vanishing components of $Y-X V^{-1}W$, and we need another condition, which corresponds to the uniqueness condition~(\ref{DC2}) in our case, where the maximally-degenerate condition is applied. As far as we investigated, we do not deny the possibility that Lagrange multipliers that seem to conflict with the condition would be fortunately determined by the consistency conditions of the secondary and the following constraints and eventually give an unique time evolution, though such an example would be difficult to achieve.}  
Thus, the condition,
\begin{align}
 Y-X V^{-1} W=0 \ ,
\label{B7}
\end{align}
would be required for a healthy constrained system, where the dynamics is uniquely solved when we specify a set of initial conditions.
Then, $\bar{\lambda}$ are completely free, and we have at most $m'-r$ secondary constraints from
\begin{align}
 -B+X V^{-1}A ~\approx~ 0 \ .
\end{align}
If we would like to determine the remaining Lagrange multipliers
$\lambda$ uniquely, we need $W=0$ as seen from \eqref{det_multiplier_upper},
which means $Y=0$ from \eqref{B7}. In addition, since $X=W^T$, $X$ also vanishes. Then,
the (possible) secondary constraints reduce to $B ~\approx~ 0$. We conclude
that we need $W=0$, $X=0$ and $Y=0$ to have the unique time development
of the system, that is, to obtain (possible) secondary constraints and
to determine the Lagrange multipliers properly.

\subsection{A simple example}
We give one of the simplest examples to understand the above discussion more concretely,
\begin{align}
 L(\theta,\dot{\theta})=i\theta_1\dot{\theta}_1+\dot{\theta_1}\theta_1\theta_2\theta_3 \ .
\label{Lagrangian_ExC}
\end{align}
The primary constraints are
\begin{align}
 \phi_1=\pi_1+i\theta_1-\theta_1\theta_2\theta_3\ , \quad 
 \phi_2=\pi_2 \ , \quad \phi_3=\pi_3 \ .
\end{align}
As easily verified with the constraints, the Hamiltonian vanishes, $H=0$, which implies that no further constraints appear from the consistency conditions, ${\dot \phi}_\alpha \approx 0$, which are concretely written as
\begin{align}
 \begin{pmatrix}
  -2(i-\theta_2\theta_3) & -\theta_1\theta_3 & \theta_1\theta_2 \\
  -\theta_1\theta_3 & 0 & 0 \\
  \theta_1\theta_2 & 0 & 0
 \end{pmatrix}
 \begin{pmatrix}
  \lambda_1\\
  \lambda_2\\
  \lambda_3
 \end{pmatrix}
 = 0 \ .
\label{C_example_consistency}
\end{align}
The second and third lines cannot be solved for $\lambda_2$ and
$\lambda_3$. Then, even if we solve the first line for
$\lambda_1$, we cannot determine $\lambda_1$ uniquely because of the dependence
on $\lambda_2$ and $\lambda_3$. It should be noticed that, since no
secondary constraints appear, the regularity condition
itself cannot rule out this example. In Lagrangian formulation, the
Euler-Lagrange equations are
\begin{align}
 \begin{pmatrix}
  -2(i-\theta_2\theta_3) & -\theta_1\theta_3 & \theta_1\theta_2 \\
  -\theta_1\theta_3 & 0 & 0 \\
  \theta_1\theta_2 & 0 & 0
 \end{pmatrix}
 \begin{pmatrix}
  \dot{\theta}_1\\
  \dot{\theta}_2\\
  \dot{\theta}_3
 \end{pmatrix}
 = 0 \ ,
\end{align}
where we see exactly the same structure with \eqref{C_example_consistency}. 
The velocities $\dot{\theta}_2$ and $\dot{\theta}_3$ are not determined, 
but we need their information to follow the time evolution of $\theta_1$. 
As a result, there is no deterministic dynamics in this system. 
The above example corresponds to $W\neq 0$, $X\neq 0$ and $Y=0$ case, 
but $W=0$, $X=0$ and $Y\neq 0$ case is easily found by turning off the first term in \eqref{Lagrangian_ExC}, giving us a similar result.

\section{Equivalence of the existence of $N$ primary constraints and the maximally-degenerate condition}
\label{Appendix2}
We have adopted $D - \C A^{-1} \B = 0$ for any $N$.
With this assumption, we have $N$ relations determining the variation of $\pi_\alpha$,
\ba
\delta \pi_\alpha - \C_{\alpha i}
A^{ij} \delta p_j 
+\left( {\cal C}_{\alpha i}A^{ij} L_{\dot{q}^j q^k} - L_{\dot{\theta}^\alpha q^k} \right) \delta q^k 
-\left( {\cal C}_{\alpha i}A^{ij} L_{\dot{q}^j \theta^\beta} +
L_{\dot{\theta}^\alpha \theta^\beta} \right) \delta \theta^\beta
= 0 \ .
\ea
In the following, we show that the condition, $D-\C A^{-1}\B=0$, indeed yields $N$ primary constraints, determining $\pi_\alpha$ in terms of the other canonical variables, and confirm the equivalence between them without relying on the integrable condition.
The variations of the Lagrangian $L(q,\dot{q},\theta,\dot{\theta})$ with respect to $\dot{\theta}^\alpha$ give
\begin{align}
	\pi_\alpha=F_\alpha(q,\dot{q},\theta,\dot{\theta}) \ , \qquad {\rm where} \quad
	F_\alpha(q,\dot{q},\theta,\dot{\theta})=\left. \frac{\partial
		L(q,\dot{q},\theta,\dot{\theta})}{\partial
		\dot{\theta}^\alpha}\right|_{q, \theta,\dot{q}} \ . 
	\label{TildePhi}
\end{align}
Similarly, the variations of the Lagrangian with respect to $\dot{q}^i$ give
\begin{align}
	p_i=G_i(q,\dot{q},\theta, \dot{\theta}) \ , \qquad {\rm where} \quad  G_i(q,\dot{q},\theta, \dot{\theta})=\left. \frac{\partial L(q,\dot{q},\theta,\dot{\theta})}{\partial \dot{q}^i}\right|_{q, \theta, \dot{\theta}} \ .
	\label{PiGi}
\end{align}
Since the derivative of these with respect to $\dot{q}^j$ coincides with the invertible matrix $A_{ij}$ defined in the text,
the inverse function theorem implies that we can locally write down
\begin{align}
	\dot{q}^i=g^i(q,p, \theta,\dot{\theta}) \ ,
\end{align}
where $g^i$ are functions. If we substitute them into \eqref{TildePhi}, we have
\begin{align}
	\pi_\alpha=F_\alpha(q,g(q,p,\theta,\dot{\theta}),\theta, \dot{\theta}) \ ,
\end{align}
and the variations with respect to $\dot{\theta}^\beta$ with keeping $q$, $\theta$ and $p$ fixed are 
\begin{align}
	\left.\frac{\partial \pi_\alpha}{\partial \dot{\theta}^\beta}\right|_{q,\theta, p}
	=\left.\frac{\partial F_\alpha}{\partial \dot{\theta}^\beta}\right|_{q, \theta, \dot{q}}
	+\left.\frac{\partial g^i}{\partial \dot{\theta}^\beta}\right|_{q, \theta, p}
	\left.\frac{\partial F_\alpha}{\partial \dot{q}^i}\right|_{q, \theta, \dot{\theta}}
	=D_{\alpha\beta}-\C_{\alpha i}A^{ij}\B_{j\beta} \ .
\end{align}
We find that the maximally-degenerate condition, $D-\C A^{-1} \B=0$, is exactly the same with the independence of $\pi_\alpha$ from $\tnd^\beta$. Therefore, under (\ref{DC1}), we actually have $N$ primary constraints~\eqref{primary},
\begin{align}
	\phi_\alpha=\pi_\alpha-F_\alpha(q,p,\theta)=0 \ .
\end{align}

\section{Dirac brackets in the maximally-degenerate case}
\label{Appendix4}
We explicitly show all the Dirac brackets between the canonical variables calculated in the maximally-degenerate case. They are
\begin{align}
 \{\theta^\alpha, \theta^\beta\}_D=-(C^{-1})^{\alpha\beta}\ ,  \qquad 
 \{\theta^\alpha, q^i\}_D=(C^{-1})^{\alpha\gamma}\frac{\partial F_\gamma}{\partial p_i}\ ,  \qquad 
 \{\theta^\alpha, p_i\}_D=-(C^{-1})^{\alpha\gamma}\frac{\partial F_\gamma}{\partial q^i}\ ,
\end{align}
\begin{align}
 \{q^i, q^j\}_D=\frac{\partial F_\alpha}{\partial p_i}(C^{-1})^{\alpha\beta}\frac{\partial F_\beta}{\partial p_j}\ ,  \qquad 
 \{q^i, p_j\}_D=\delta^i_{~j}-\frac{\partial F_\alpha}{\partial p_i}(C^{-1})^{\alpha\beta}\frac{\partial F_\beta}{\partial q^j}\ ,  \qquad 
 \{p_i, p_j\}_D=\frac{\partial F_\alpha}{\partial q^i}(C^{-1})^{\alpha\beta}\frac{\partial F_\beta}{\partial q^j}\ ,  
\end{align}
and
\begin{align}
 \{\theta^\alpha, \pi_\beta\}_D=-\delta^\alpha_{~\beta}+(C^{-1})^{\alpha\gamma}\frac{\partial F_\gamma}{\partial \theta^\beta} \ , \qquad 
 \{\pi_\alpha, \pi_\beta\}_D=-\frac{\partial F_\gamma}{\partial \theta^\alpha}(C^{-1})^{\gamma\delta}\frac{\partial F_\delta}{\partial \theta^\beta} \ , 
\end{align}
which are found to be consistent with \eqref{dirac_pi_F} by taking into account \eqref{PC} and the following identities,
\begin{align}
 \{\theta^\alpha, F_\beta\}_D=\{\theta^\alpha,\theta^\gamma\}_D\frac{\partial F_\beta}{\partial \theta^\gamma}+\{\theta^\alpha,q^i\}_D\frac{\partial F_\beta}{\partial q^i}+\{\theta^\alpha,p_i\}_D\frac{\partial F_\beta}{\partial p_i} \ ,
\end{align}
\begin{align}
 \{F_\alpha, F_\beta\}_D
=&\{\theta^\gamma,\theta^\delta\}_D\frac{\partial F_\alpha}{\partial \theta^\gamma}\frac{\partial F_\beta}{\partial \theta^\delta}
+\{\theta^\gamma,q^i\}_D\Bigl(\frac{\partial F_\alpha}{\partial q^i}\frac{\partial F_\beta}{\partial \theta^\gamma}+\frac{\partial F_\alpha}{\partial \theta^\gamma}\frac{\partial F_\beta}{\partial q^i}\Bigr)
+\{\theta^\gamma,p_i\}_D\Bigl(\frac{\partial F_\alpha}{\partial p_i}\frac{\partial F_\beta}{\partial \theta^\gamma}+\frac{\partial F_\alpha}{\partial \theta^\gamma}\frac{\partial F_\beta}{\partial p_i}\Bigr) \nonumber\\
&+\{q^i,q^j\}_D\frac{\partial F_\alpha}{\partial q^i}\frac{\partial F_\beta}{\partial q^j}
+\{p_i,p_j\}_D\frac{\partial F_\alpha}{\partial p_i}\frac{\partial F_\beta}{\partial p_j}
+\{q^i,p_j\}_D\Bigl(\frac{\partial F_\alpha}{\partial q^i}\frac{\partial F_\beta}{\partial p_j}-\frac{\partial F_\alpha}{\partial p_j}\frac{\partial F_\beta}{\partial q^i}\Bigr) \ .
\end{align}

\section{An example of ghost free boson-fermion system in field theory}
\label{Appendix5}
In this Appendix, we give a simple extension to a boson-fermion system in the context of field theory. 
Let us introduce a real scalar field $\phi(t, \bx)$ and a Weyl fermion $\psi^\alpha(t, \bx)$ $(\alpha=1, 2)$, 
and consider the following Lagrangian density~\footnote{Please note that upper indices are lowered by $\epsilon_{\alpha\beta}$ instead of $\delta_{\alpha\beta}$, which we used in the text and in other appendices, i.e., $\psi_\alpha=\epsilon_{\alpha\beta}\psi^\alpha$ rather than $\theta_\alpha=\delta_{\alpha\beta}\theta^\beta$.},
\ba
\Lag &=& {1\over 2} (\p_\mu \phi)^2 
-i \p^\mu \phi (\psi^\alpha \p_\mu \psi_\alpha- \psib_{\alphad} \p_\mu \psib^\alphad)
+ {i\over2} \left(
\psib^\alphad \sigma_{\alpha\alphad}^\mu \p_\mu \psi^\alpha 
-\p_\mu \psib^\alphad \sigma_{\alpha\alphad}^\mu \psi^\alpha
\right)\nonumber\\
&&-{1\over 2}(\psi^\alpha \p_\mu \psi_\alpha)^2 + (\psi^\alpha \p_\mu \psi_\alpha)(\psib_\alphad \p^\mu \psib^\alphad)-{1\over 2}(\psib_\alphad \p_\mu \psib^\alphad)^2 \ .
\label{LagFT}
\ea
Hereafter, we follow the spinor conventions in \cite{WB:1992} and the metric signature convention, $(+, -, -, -)$.
The canonical momenta are given by
\ba
\pi_\phi^\mu &=& {\p \Lag \over \p (\p_\mu \phi)} = \p^\mu \phi -i (\psi^\alpha \p^\mu \psi_\alpha -\psib_\alphad \p^\mu \psib^\alphad)\ ,\\
\pi_{\psi^\alpha}^\mu &=& {\p \Lag \over \p (\p_\mu \psi^\alpha)} 
= -i \p^\mu \phi\, \psi_\alpha 
-{i\over2} \psib^\alphad \sigma_{\alpha\alphad}^\mu 
- \psi_\alpha (\psi^\beta \p^\mu \psi_\beta) + \psi_\alpha (\psib_\alphad \p^\mu \psib^\alphad)  = -i \pi_\phi^\mu \psi_\alpha-{i\over2} \psib^\alphad \sigma_{\alpha\alphad}^\mu \ ,
\label{Pri1}\\
\pi_{\psib^\alphad}^\mu &=& {\p \Lag \over \p (\p_\mu \psib^\alphad)} 
= -i \p^\mu \phi \,\psib_\alphad 
-{i\over2} \sigma_{\alpha\alphad}^\mu \psi^\alpha
+ \psib_\alphad (\psib_\betad \p^\mu \psib^\betad)  
- \psib_\alphad (\psi^\beta \p^\mu \psi_\beta)= -i \pi_\phi^\mu \psib_\alphad-{i\over2} \sigma_{\alpha\alphad}^\mu \psi^\alpha\ ,
\label{Pri2}
\ea
where we have eliminated $\p^\mu \phi$ by using $\pi_\phi^\mu$ 
in the second equalities of the expression of $\pi_{\psi^\alpha}^\mu $ and $\pi_{\psib^\alphad}^\mu$. 
Note that the momenta for $\psi$ and $\psib$ are related through the anti-Hermitian relation, $(\pi_{\psi^\alpha}^\mu)^\dagger = -\pi_{\psib^\alphad}^\mu$.
As one can see from (\ref{Pri1}) and (\ref{Pri2}), the fermionic momenta are functions of other canonical variables,
just as discussed in Sec.~\ref{Subsection:DegeneracyConditions}. 
Therefore, the zero-th components of the canonical momenta for the fermion yield four primary constraints,
\ba
\Phi_{\psi^\alpha} &\equiv&\pi_{\psi^\alpha}^0+i \pi_\phi^0 \psi_\alpha 
+{i\over2} \psib^\alphad \sigma_{\alpha\alphad}^0 =0\ ,\label{PRI1}\\
\Phi_{\psib^\alphad} &\equiv& \pi_{\psib^\alphad}^0+ i \pi_\phi^0 \psib_\alphad 
+{i\over2} \sigma_{\alpha\alphad}^0 \psi^\alpha=0\ .\label{PRI2}
\ea
We define Hamiltonian and total Hamiltonian as 
\ba
H = \int d^3 \bx\; {\cal H} \ , \qquad
H_T = \int d^3 \bx\; {\cal H}_T \ ,
\ea
where their densities are
\begin{align}
{\cal H} = \dot{\phi}\pi_\phi^0+\dot{\psi}^\alpha\pi_{\psi^\alpha}^0+\dot{\bar{\psi}}^{\dot{\alpha}}\pi_{\bar{\psi}^{\dot{\alpha}}}^0-{\cal L}\ , \qquad 
{\cal H}_T ={\cal H}+ \Phi_{\psi^\alpha} \lambda^\alpha + \Phi_{\psib^\alphad} {\bar \lambda}^\alphad\ .
\end{align}
Now we would like to use the Poisson bracket, defined as
\begin{align}
 &\{\F (t, \bx), \G(t, \by) \}\\
 &~~= \int d^3 z \Biggl[
{\delta \F(t, \bx)  \over \delta \phi (t, \bz) }{\delta \G(t, \by)  \over \delta \pi_\phi^0(t, \bz) }
-{\delta \F(t, \bx)  \over  \delta \pi_\phi^0(t, \bz) }{\delta \G(t, \by)  \over \delta \phi (t, \bz)}\nonumber\\
 &~~~~~~~~~\qquad~
+(-)^{\varepsilon_\F} 
\Biggl(
{\delta \F(t, \bx)  \over \delta \psi^\alpha (t, \bz) }{\delta \G(t, \by)  \over \delta \pi_{\psi^\alpha}^0(t, \bz) }
+{\delta \F(t, \bx)  \over \delta \pi_{\psi^\alpha}^0(t, \bz) }{\delta \G(t, \by)  \over \delta  \psi^\alpha (t, \bz) }
+{\delta \F(t, \bx)  \over \delta \psib^\alphad (t, \bz) }{\delta \G(t, \by)  \over \delta \pi_{\psib^\alphad}^0(t, \bz) }
+{\delta \F(t, \bx)  \over \delta  \pi_{\psib^\alphad}^0(t, \bz) }{\delta \G(t, \by)  \over \delta\psib^\alphad (t, \bz)}
\Biggr)
\Biggr] \ .~~~~~~\nonumber\\ 
\end{align}
The Poisson brackets between the canonical variables are given by
\ba
\{\phi (t, \bx), \,\pi_\phi^0 (t, \by) \} &=& \delta^3(\bx -\by) \ ,\\
\{\psi^\alpha (t, \bx), \, \pi_{\psi^\beta}^0 (t, \by) \} &=& -\delta^{\alpha}_{~\beta}\, \delta^3(\bx -\by) \ ,\\
\{\psib^\alphad (t, \bx), \,\pi_{\psib^\betad}^0 (t, \by)\} &=& -\delta^{\alphad}_{~\betad}\, \delta^3(\bx -\by)\ ,
\ea
while the other Poisson brackets vanish. 
Then, the Poisson brackets between the primary constraints are 
\ba
\{\Phi_{\psi^\alpha} (t, \bx), \,\Phi_{\psi^\beta}  (t, \by)\} &=& 0\ , \\
\{\Phi_{\psib^\alphad} (t, \bx), \,\Phi_{\psib^\betad} (t, \by) \} &=& 0 \ ,\\
\{\Phi_{\psi^\alpha} (t, \bx), \,\Phi_{\psib^\alphad}  (t, \by) \} &=& -i \sigma_{\alpha\alphad}^0\delta^3(\bx -\by)\ . 
\ea
 Then the time-evolution of the primary constraints are 
 \ba
 {\dot \Phi}_{\psi^\alpha} (t, \bx)&=& \{ \Phi_{\psi^\alpha}(t, \bx) , \, H_T\} 
 = \{ \Phi_{\psi^\alpha} (t, \bx), \, H\}
 +\int d^3 \by \,\{ \Phi_{\psi^\alpha}(t, \bx), \,   \Phi_{\psib^\alphad}(t, \by)\}{\bar \lambda}^\alphad(t, \by)\approx 0 \ ,\\
 {\dot \Phi}_{\psib^\alphad}(t, \bx)&=& \{ \Phi_{\psib^\alphad}(t, \bx), \, H_T \} 
 = \{ \Phi_{\psib^\alphad}(t, \bx) , \, H\}
 +\int d^3 \by \,\{\Phi_{\psib^\alphad}(t, \bx), \,  \Phi_{\psi^\alpha}(t, \by)\}\lambda^\alpha(t, \by)  \approx 0 \ .
 \ea
Thus, all the Lagrange multipliers are fixed, 
\begin{align}
 \begin{pmatrix}
  \lambda^\alpha(t,\bx)\\
  \bar{\lambda}^{\dot{\alpha}}(t,\bx)
 \end{pmatrix}
 =-i
 \begin{pmatrix}
  \bar{\sigma}^{0 \dot{\alpha}\alpha}\{\Phi_{\bar{\psi}^{\dot{\alpha}}}(t,\bx),H\}\\
  \bar{\sigma}^{0 \dot{\alpha}\alpha}\{\Phi_{\psi^{\alpha}}(t,\bx),H\}
 \end{pmatrix}\,,
\end{align}
and the primary constraints  (\ref{PRI1}) and (\ref{PRI2}) are 
second-class. Thus, the number of degrees of freedom is 
\ba
\textrm{Degrees of freedom} = {2 \times 1 \,\textrm{(bosonic)} + 2 \times 4  \,\textrm{(fermionic)} - 4  \,\textrm{(constraints)} \over 2} 
= 1 \,\textrm{(bosonic)}+2 \,\textrm{(fermionic)} \ , \nonumber\\
\ea
as desired. 
Therefore, the theory (\ref{LagFT}) is  free of fermionic ghosts. 

Let us finally check the consistency with equations of motion derived in Lagrangian formulation. 
The equation of motion for $\phi$ is given by
\ba
E_\phi &\equiv& \square \phi - i \p_\mu (\psi^\alpha \p^\mu \psi_\alpha -\psib_\alphad \p^\mu \psib^\alphad) =0 \ ,
\label{phi_eom_org}
\ea
and those for $\psi^\alpha$ and $\psib^\alphad$ 
can be written as 
\ba
i \p_\mu \psib^\alphad \sigma_{\alpha\alphad}^\mu + i E_\phi \psi_\alpha&=&0 \ ,\\
i  \sigma_{\alpha\alphad}^\mu \p_\mu \psi^\alpha+ i E_\phi \psib_\alphad&=&0 \ .
\ea
The second terms in both equations, which contain the second derivatives of the fermion, 
vanish after using the equation of motion for $\phi$, 
and we then have the familiar Weyl equations, which results in 
\begin{align}
 \square \psi^\alpha=0 \ , \qquad \square \bar{\psi}^{\dot{\alpha}}=0 \ ,
\end{align}
by making use of the properties of the sigma matrix.
We then substitute them back into \eqref{phi_eom_org} to have
\begin{align}
 \square \phi - i \partial_\mu \psi^\alpha \p^\mu \psi_\alpha +i \partial_\mu \psib_\alphad \p^\mu \psib^\alphad=0\ ,
\end{align}
and this is nothing but the second-order differential equation for $\phi$,
which ensures that the number of degrees of freedom including fermionic degrees of freedom is $3$. 


\end{document}